\documentclass[apj, twocolumn, numberedappendix, appendixfloats]{openjournal}

\usepackage{lipsum}

\usepackage{xcolor}
\usepackage{textgreek}
\usepackage[utf8]{inputenc}
\usepackage[english]{babel}

\usepackage{hyperref}
\hypersetup{
    unicode, 
    colorlinks=true,
    linkcolor=linkcolor,
    citecolor=linkcolor,
    filecolor=linkcolor,
    urlcolor=linkcolor,
}
\usepackage{color,colortbl}
\definecolor{linkcolor}{rgb}{0.0,0.3,0.5}
\usepackage{tensind}
\tensordelimiter{?}
\DeclareGraphicsExtensions{.bmp,.png,.jpg,.pdf}
\usepackage{verbatim}
\usepackage[normalem]{ulem}
\usepackage{orcidlink}
\usepackage{soul}

\usepackage{amsmath}
\usepackage{amssymb}

\DeclareRobustCommand{\ion}[2]{%
\relax\ifmmode
\ifx\testbx\f@series
{\mathbf{#1\,\mathsc{#2}}}\else
{\mathrm{#1\,\mathsc{#2}}}\fi
\else\textup{#1\,{\mdseries\textsc{#2}}}%
\fi}

\usepackage{graphicx}	
\usepackage{subfigure}
\usepackage{rotating}
\usepackage{longtable}

\graphicspath{ {./figures/} }

\shortauthors{Y.~A. Gordon et al.}
\shorttitle{IceCube/VLASS Cross Associations}

\begin{document}
\title{New Signs Pointing Toward a Correlation Between Astrophysical Neutrinos\\and Radio Flares
\vspace{-4em}
}

\author{Yjan~A. Gordon$^{1, 4}$\orcidlink{0000-0003-1432-253X}}
\author{Peter~S. Ferguson$^{2}$\orcidlink{0000-0001-6957-1627}}
\author{Eric~J. Hooper$^{3}$\orcidlink{0000-0003-0713-3300}}
\author{Michael~N. Martinez$^{1, 5}$\orcidlink{0000-0002-8397-8412}
\vspace{0.5em}}

\thanks{Corresponding Author: Yjan~A. Gordon \\ \href{mailto:yjan.a.gordon@gmail.com}{yjan.a.gordon@gmail.com}}

\affiliation{$^1$Department of Physics, University of Wisconsin-Madison, 
1150 University Ave., Madison, WI 53706, USA\\
$^2$Department of Astronomy, University of Washington, 3910 15th Ave. NE, Seattle, WA 98195, USA\\
$^3$Department of Astronomy, University of Wisconsin-Madison, 
475 N. Charter St., Madison, WI 53706, USA\\
$^4$Department of Physics, University of Wisconsin-Whitewater, 800 W. Main St., Whitewater, WI 53190, USA\\
$^5$Department of Physics, University of California, Merced, 5200 N Lake Road, Merced, CA 95341, USA}

\begin{abstract}
    We use data from the first two epochs of the Very Large Array Sky Survey (VLASS) and the IceCube Neutrino Observatory to search for evidence of a correlation between radio variability and the detection of astrophysical neutrinos.
    Our work probes the radio population down to $S_{3\,\text{GHz}}\gtrsim$ a few mJy, more than an order of magnitude fainter than previous analyses that aim to cross correlate radio variability with neutrino detections.
    We find an excess number of associations between flaring radio sources and neutrinos that were detected between the first and second VLASS observations at $>2\sigma$ confidence.
    This excess is consistent with radio flares contributing $\sim13\,\%$ of the astrophysical neutrinos observed by IceCube.
    Notably, $\sim90\,\%$ of the radio flares associated with neutrinos are not detected at either $\gamma$-ray or X-ray wavelengths,
    highlighting the importance of deep radio observations for identifying potential electromagnetic counterparts to astrophysical neutrinos.
    No excess in the number of associations between the wider radio-variable population and the IceCube neutrinos is seen when no time constraint is placed on the neutrino detection.
    We predict that data from future VLASS epochs will see an excess number of associations between radio flares and neutrinos at the $>3\sigma$ level, and expected improvements to the positional constraints on the neutrinos may increase that confidence to $>5\sigma$, should our results be representative.\\
\end{abstract}

\begin{keywords}
{
\href{http://astrothesaurus.org/uat/16}{Active galactic nuclei (16)},
\href{http://astrothesaurus.org/uat/164}{Blazars (164)},
\href{http://astrothesaurus.org/uat/508}{Extragalactic radio sources (508)},
\href{http://astrothesaurus.org/uat/1100}{Neutrino Astronomy (1100)},
\href{http://astrothesaurus.org/uat/1338}{Radio Astronomy (1338)},
\href{http://astrothesaurus.org/uat/2109}{Time Domain Astronomy (2109)}
}
\end{keywords}

\maketitle


\section{Introduction} \label{sec:intro}

Astrophysical neutrinos present a unique way to study the cosmos.
Their low interaction cross section allows them to pass through highly dense environments that are opaque to electromagnetic radiation (particularly at shorter wavelengths).
Moreover, the extremely low probability of interaction means that neutrinos can reach Earth from the farthest and earliest reaches of the Universe\textemdash unlike light, neutrinos have no effective horizon.
In recent years, the IceCube Neutrino Observatory has confirmed the existence of high energy (HE; $E_\nu\gtrsim 1\,$TeV) astrophysical neutrinos \citep{IceCube2013}. 
Some of these HE neutrinos are diffuse emission from the plane of the Milky Way \citep[][]{IceCube2023-milkyway}, potentially resulting from cosmic rays in the Milky Way's interstellar medium.
Further to this, two extragalactic point sources of HE neutrinos have been identified so far: the blazar \text{TXS $0506+056$}, and the Seyfert galaxy \text{NGC $1068$} \citep{IceCube2018-txs0506, IceCube2022-ngc1068}.
Active galactic nuclei (AGN) such as \text{TXS $0506+056$} and \text{NGC $1068$} are intriguing candidates as neutrino sources, with their central engines potentially hosting a number of mechanisms that should in theory produce HE neutrinos, e.g., magnetic reconnection events in the accretion disk halo, or shocks in the jet or accretion disk \citep{Mannheim1995, Bednarek1999, Khiali2016, Blandford2019, Murase2023}.
However, with only a handful of neutrino sources identified to date, it is difficult to determine what processes are responsible for the HE astrophysical neutrinos we observe.
The relativistic outflows and radiative processes found in a variety of energetic astronomical phenomena may result in the particle acceleration necessary for HE neutrino production.
Consequently, events such as gamma-ray bursts (GRBs), supernovae (SNe), and tidal disruption events (TDEs) provide theoretical alternative sources for observed HE neutrinos \citep{Razzaque2003, Denton2018, Morgan2019, Stein2021, Winter2024}.

Identifying electromagnetic counterparts to neutrino detections has proven difficult.
Any one HE neutrino detection has a relatively small chance of being a real astrophysical neutrino.
Moreover, the typically large positional uncertainties for neutrino detections encompass many potential sources of origin.
A representative circular positional error for an IceCube detection has a radius of $\sim1^{\circ}$ \citep{Aartsen2017} covering an area of $\sim 3\,\text{deg}^{2}$.
Modern deep optical surveys typically detect hundreds of thousands of sources per square degree; for example, the on-sky source density, $\rho$, of the Hyper Suprime Cam wide survey is $\sim270,000\,\text{deg}^{-2}$ \citep{Aihara2022}.
It is therefore usually not possible to confidently associate a single neutrino detection with an electromagnetic counterpart.
As a result, observational studies of the origins of astrophysical neutrinos typically rely on one of two approaches.
First, associations of multiple neutrino events with a single point of origin increase the likelihood that the source in question is a neutrino emitter.
This is the approach that led to \text{TXS $0506+056$} and \text{NGC $1068$} being identified as neutrino-producing AGN \citep{IceCube2018-txs0506, IceCube2022-ngc1068}.
Second, so-called stacking experiments that combine large numbers of low-probability associations in order to identify trends in whether certain source types are more often associated with neutrinos than would be expected from random matching of two unrelated populations.
This latter approach cross correlates multiple neutrino detections with multiple objects of a given type, e.g., star-forming galaxies or blazars \citep[][]{Bechtol2017, Aartsen2017_blazars, Fang2020, Abbasi2024}. 

The production mechanism for neutrinos invariably results in high energy electromagnetic radiation.
Consequently, incorporating time-domain information into searches for the electromagnetic counterparts to HE neutrinos can reduce uncertainty in the associations between individual neutrino events and potential sources of origin. 
For instance, neutrinos have been observed at similar times to $\gamma$-ray, X-ray and radio outbursts in blazars that are spatially coincident with the neutrino detection \citep[e.g.,][]{Kadler2016, KM3blazar}.
Stacking experiments that search for correlations between electromagnetic flares and HE neutrinos have shown some intriguing results.
For example, \citet{Plavin2020}, \citet{Hovatta2021} and \citet{Kouch2024} all find a potential correlation between radio outbursts in AGN and HE neutrinos at the $\sim2\,\sigma$ level; \citet{Lu2025} identified a likely correlation between optical flares associated with TDEs and HE neutrinos at $\sim2.5\,\sigma$ confidence; while \citet{vanVelzen2024} report a $3.6\,\sigma$ confidence association between HE neutrinos and what they term \emph{``accretion flares''} (a superset of optical outbursts in AGN and TDEs), using a combination of optical and infrared observations.

Radio observations are a particularly promising avenue in the search for electromagnetic counterparts to HE astrophysical neutrinos for three key reasons.
First, many of the phenomena thought to produce HE neutrinos (AGN, SNe, GRBs, and TDEs) are expected to emit radio waves.
Indeed, the two confirmed neutrino point sources to date are both bright at radio frequencies \citep[e.g.,][]{Kuehr1981, LaurentMuehleisen1997}.
Second, the long wavelengths of radio band emission are less easily obscured than shorter wavelength radiation.
Third, radio sources often have higher redshifts than objects not detected in the radio, and thus sample a larger volume of the Universe \citep{Gordon2025}. 
As such, for neutrinos originating at high redshift, the radio emission (if produced) may be bright enough to be detected when the source is too faint to (currently) be observed at other wavelengths, potentially increasing the number of neutrinos for which an electromagnetic counterpart is detected.

The advent in recent years of deep and wide-area time-domain radio sky surveys, such as the Karl G. Jansky Very Large Array Sky Survey \citep[VLASS,][]{Lacy2020} and the Variables And Slow Transients survey \citep[VAST,][]{Murphy2013}, opens up new possibilities in searching for electromagnetic counterparts to HE neutrinos \citep{Perger2024, Perger2025, Filipovic2025}.
Previously, searches for correlations between radio variability and HE neutrino detections have been limited to bright sources with radio flux densities, $S$, typically brighter than a few hundred mJy.
In contrast, the latest generation of radio surveys can detect variability in objects that are orders of magnitude fainter, with $S\gtrsim$ a few mJy \citep{Gordon2025}, detecting potential neutrino counterparts that are intrinsically less luminous and/or across a larger cosmic volume.
To that end, in this work we search for evidence of a correlation between variable radio sources detected in VLASS and HE neutrinos detected by IceCube.

The rest of this paper is laid out as follows.
Section \ref{sec:data} describes the VLASS and IceCube data used in this work.
In Section \ref{sec:xcorrelation} we describe the process of associating variable radio sources with neutrino events and details of the statistical analysis we perform on the identified radio-neutrino associations.
We discuss our results along with future prospects in Section \ref{sec:discussion}, and a summary of this work is given in Section \ref{sec:summary}. 
The catalog of associations between flaring radio sources and HE neutrinos that we produce in this work is made available to the community, as described in Appendix \ref{ap:assocdata}.

\section{Data} \label{sec:data}

\subsection{VLASS}

The Karl G. Jansky Very Large Array Sky Survey \citep[VLASS,][]{Lacy2020}, is a multi-epoch survey of the entire sky north of $\delta =-40^{\circ}$ at $\nu \sim 3\,$GHz that began in 2017.
The survey has a synthesized beam size of $\sim3\arcsec$ and a typical rms noise of $130\,\mu\text{Jy}\,\text{beam}^{-1}$ in each epoch, making VLASS the highest resolution, deep, near-all-sky radio continuum survey to date.
Approximately $1.8\times10^{6}$ sources brighter than $\sim 1\,$mJy are detected in a VLASS epoch \citep{Gordon2021}, with $\sim2.5\,$years between VLASS epochs.
Four VLASS epochs were observed between 2017 and 2026 \citep{Lacy2020, Nyland2023}, with full catalogs for the first two epochs available at the time of writing.

We identified $3,618$ variable radio sources in \citet{Gordon2025} based on Epoch 1 (September 2017 - July 2019) and Epoch 2 (June 2020 - February 2022) of VLASS.
The majority of these sources were found to be probable blazars, and using the data from \citet{Gordon2025} not only allows us to know where they are on the sky but also provides a window into when some of these sources experienced outbursts of activity.
In \citet{Gordon2025} we characterized the variable radio population for sources varying in flux density by $30\,\%$ or more across the two epochs.
Here we are specifically looking for more substantive changes in brightness and select only those variables exhibiting at least a $50\,$\% change in flux density. 
Explicitly, we select the $1,928$ sources with $|m| > 0.4$ for use in this work, where $m = \Delta S/\overline{S}$ is the modulation index for a source with flux density measurements $S$ \citep[see also Section 3.3 and Table 1 in ][]{Gordon2025}.
These objects have mean brightnesses, $\mu_{S}=(S_{\text{Epoch }1} + S_{\text{Epoch }2})/2$, in the range $1\,\text{mJy} \lesssim \mu_{S}\lesssim1\,\text{Jy}$ with a median value of $\mu_{S,\ P50}\approx14\,$mJy.
Our data therefore sample a substantially fainter population than previous works that are typically limited to the radio variables in dedicated monitoring programs, generally brighter than a few hundred mJy \citep[e.g.,][]{Hovatta2021, Kouch2024}.
The majority of the radio variables in VLASS are AGN, with $\sim2/3$ having infrared colors consistent with the gamma-ray emitting blazar population, while objects such as stars, galaxies not identified as AGN, and microquasars also contribute to the radio-variable population \citep{Gordon2025}.
We perform no cuts to the radio sample on source type\textemdash we do not, for example, specifically select stars or galaxies based on multiwavelength data.
The advantage of such a source-agnostic selection is that we do not eliminate potential neutrino sources, with one of the major objectives of this work being to assess the efficacy of using faint radio time-domain sky surveys in identifying electromagnetic counterparts to astrophysical neutrinos. 

\subsection{IceCube}

The IceCube Observatory \citep{Aartsen2013, IceCube2013} detects Cherenkov radiation caused by charged leptons produced during the interactions of high energy neutrinos with the Antarctic ice.
Since 2012, IceCube has detected $\sim3,000$ muons a second, the vast majority of which are from cosmic rays \citep{Aartsen2016}.
Real astrophysical neutrinos are thought to account for around $30$ events per year in the IceCube data, and discrete tracks created by their interaction with the ice can be traced to an on-sky point of origin.
The typical angular precision of such tracks is on the order of $1^{\circ}$, depending on factors such as the neutrino energy and track length \citep{Aartsen2017}.

\begin{figure}
    \centering
    \includegraphics[trim={1cm 1cm 2cm 1}, clip, width=\columnwidth]{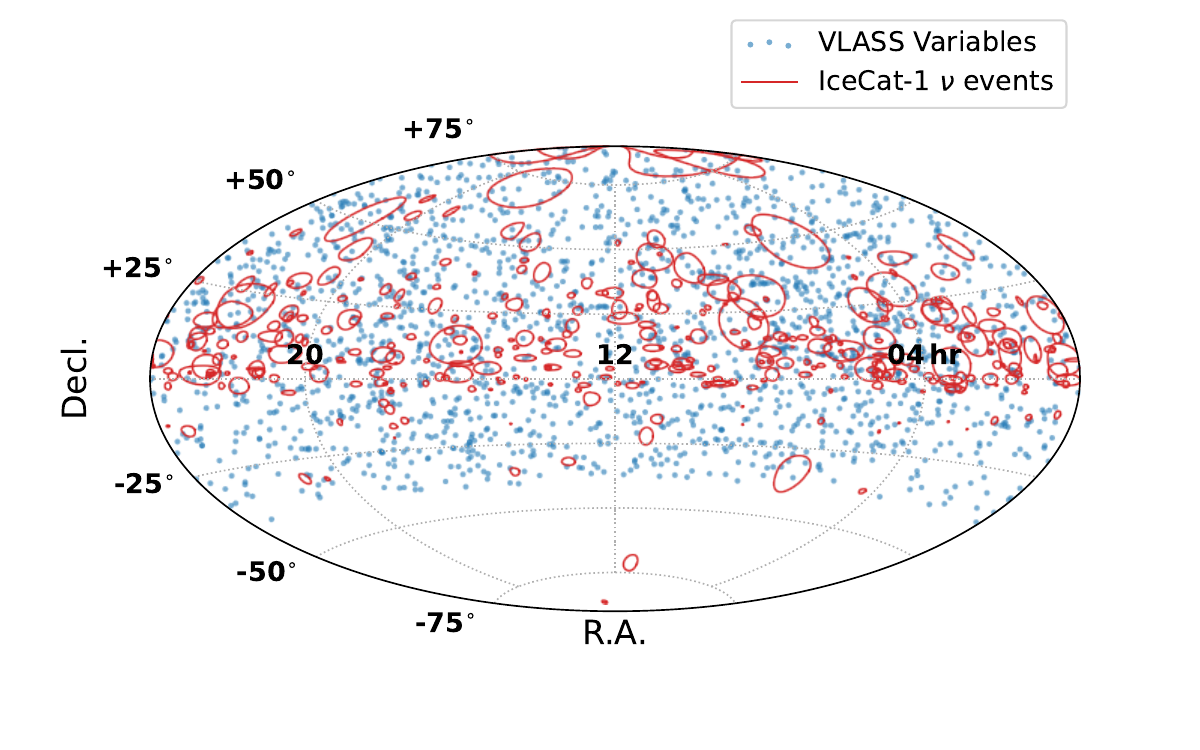}
    \caption{The on-sky distribution of VLASS variables (blue dots) and the $90\,\%$ confidence positional uncertainties of neutrino events in IceCat-1 (red lines) used in this work shown on an Aitoff projection in equatorial coordinates.}
    \label{fig:aitoff-icecube-vlass}
\end{figure}

To search for neutrino detections potentially associated with variable sources, we make use of the IceCat-1 catalog \citep{Abbasi2023}.
IceCat-1 lists track-like events detected by IceCube since 2011 that are considered likely to have an astrophysical origin.
Typically these have a $\texttt{signal}\gtrsim0.3$, where the \texttt{signal} is ostensibly a probability of being a real astrophysical neutrino that is based on the modeled neutrino energy and likely declination of origin.
Some of the events recorded in IceCat-1 have $\texttt{signal}<0.3$; these are very high energy events ($E_{\nu}>100\,$TeV) that are identified either as targets for follow up gamma-ray observations, or as a result of having extremely high energies \citep[$E_{\nu}>500\,$TeV; for full details we refer the reader to Section 2 of][]{Abbasi2023}.
In this work, we use version 2 of IceCat-1\footnote{\url{https://dataverse.harvard.edu/dataset.xhtml?persistentId=doi:10.7910/DVN/SCRUCD}}, which lists 348 IceCube track events up until October 2023.
In Figure \ref{fig:aitoff-icecube-vlass} we show how the on-sky origins of the neutrinos (red rings that enclose the $90\,\%$ positional uncertainty for each neutrino) are distributed relative to the coordinates of the variable VLASS sources (blue dots).
It is worth noting that there is a strong declination selection effect in the IceCat-1 catalog where neutrinos are easier to detect at equatorial and lower northern declinations. 

\section{Cross Association of VLASS Variables with IceCat-1} \label{sec:xcorrelation}

\subsection{Associating Radio Sources with Neutrino Events} \label{ssec:assocs}

    
    
\begin{figure*}
    \centering
    \subfigure[]{\includegraphics[trim={0 0 0.4cm 1.8cm}, clip, width=0.66\columnwidth]{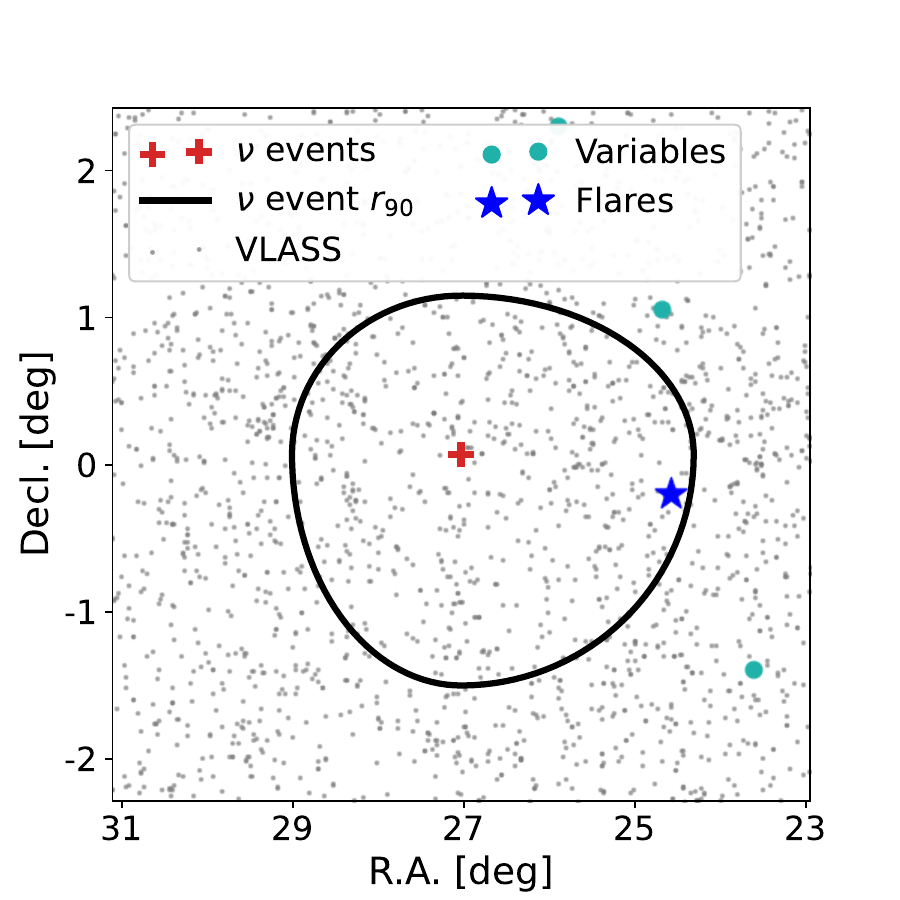}}
    \subfigure[]{\includegraphics[trim={0 0 0.4cm 1.8cm}, clip, width=0.66\columnwidth]{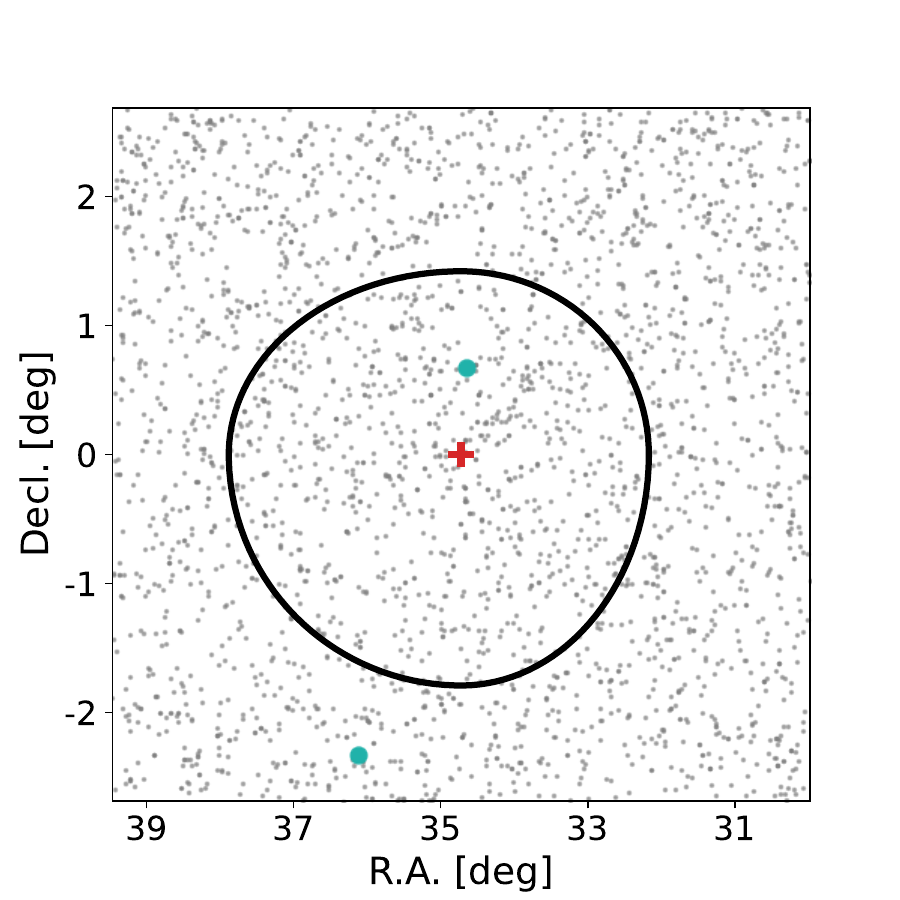}}
    \subfigure[]{\includegraphics[trim={0 0 0.4cm 1.8cm}, clip, width=0.66\columnwidth]{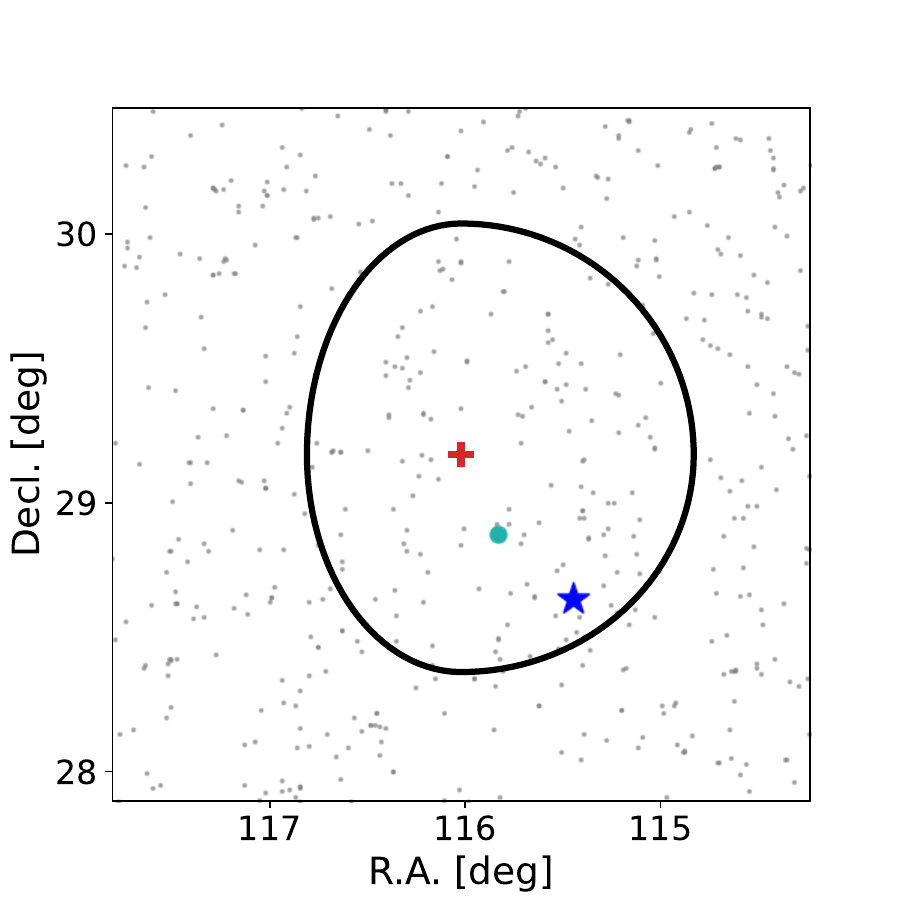}}
    \caption{Example IceCube events consistent with VLASS sources.
    In all panels the red cross shows the most likely origin coordinates of the neutrino, the black ellipse shows the $90\,$\% positional confidence, gray dots show VLASS sources, light green circles are variable VLASS sources, and blue stars highlight flares in VLASS that are consistent with the neutrino event time.
    Note that the apparent relative sparsity of VLASS sources in panel (c) is an effect of the panel being zoomed in on a neutrino event with a smaller positional uncertainty than in panels (a) and (b).
    }
    \label{fig:icecube-x-vlass}
\end{figure*}

In order to assess if a correlation exists between radio sources and HE neutrinos, we must first identify which neutrinos could be associated with which radio sources. 
The radio source positions in the sky are very well constrained, having a typical astrometric precision on the order of a few tenths of an arcsecond \citep{Gordon2021}.
The neutrino source positions have much poorer constraints, typically having positional uncertainties on the order of a degree \citep{Aartsen2017}.
IceCat-1 provides positive and negative $90\,\%$ confidence limits on the R.A. and Decl. of the neutrino detections. 
We use these uncertainties to define error ellipses in which we expect the radio counterpart (should there be one) to lie, and consider any of our VLASS sources within this ellipse to be a \textit{spatially associated} candidate for the neutrino source (e.g., see Figure \ref{fig:icecube-x-vlass}).
In addition to this spatial association, the availability of flux densities from two VLASS epochs allows us to ascertain if there is any \textit{temporal association} between the neutrino event and the radio variability. 
For instance, if a neutrino is associated with a radio flare, one might expect its observation date to be correlated with the increase in radio brightness of the source.
We consider a VLASS source to be temporally associated with a neutrino if the neutrino was detected in a time window defined by the observation dates of the first and second VLASS epochs.



We define 16 test samples of VLASS variables with which to search for evidence of a correlation with HE neutrinos.
The first of these is the full sample of $1,928$ variables, for which we search for a spatial association only.
This allows us to search for any variable radio sources that may have produced neutrinos in an outburst prior to the variability captured by VLASS between epochs 1 and 2, and we refer to this sample as the \textit{`full variable sample'} throughout the rest of this work.
For the other $15$ samples we are interested in the flaring radio sources, the $1,093$ radio variables that were brighter in Epoch 2 of VLASS than in Epoch 1 as a result of some outburst of activity between the two VLASS observations.
For these sources we are searching for both spatial \textit{and} temporal associations with the IceCat-1 neutrinos.

We define a temporal association as having a neutrino detected within a time window defined by the VLASS observation dates.
For instance, the simplest time window to consider would be a neutrino detected at a time, $t_{\nu}$, that is after the Epoch 1 observation, $t_{\text{VLASS 1}}$, and before the Epoch 2 observation, $t_{\text{VLASS 2}}$, i.e., $t_{\text{VLASS 1}} < t_{\nu} < t_{\text{VLASS 2}}$.
However, in blazars\textemdash one of the principal theorized sources of origin for HE astrophysical neutrinos\textemdash radio outbursts can lag behind $\gamma$-ray flares by $\sim$ a few hundred days \citep{MaxMoerbeck2014, Fuhrmann2014}.
Given that the neutrino production will result in $\gamma$-ray emission, one might expect the radio flare to also be delayed relative to the neutrino event.
\citet{MaxMoerbeck2014} found a median lag of $150\,$days between $\gamma$-ray and radio flares in blazars when such a time lag was detected with high confidence.
The delays reported by \citet{MaxMoerbeck2014} are based on radio observations at $15\,$GHz, a substantially higher frequency than the $3\,$GHz VLASS observations.
\citet{Fuhrmann2014} report delays between $\gamma$-ray and radio flares that are twice as long at $2.6\,$GHz than at $15\,$GHz, likely due to the impacts of synchrotron self-absorption on the opacity of the blazar jet.
We might, therefore, expect a delay of up to $\sim300\,$days between neutrino production and a flare at $\sim3\,$GHz.
Furthermore, the typical time between VLASS observations of the same object is $\sim1,000$ days, adding a smearing effect into our time correlation analysis.

In order to account for the potential delay between neutrino detection and the observed flare in VLASS, and the inherent imprecision of the VLASS cadence, we create 15 test samples based on our $1,093$ flaring VLASS sources, where the temporal association is defined as $t_{\text{VLASS 1}} < t_{\nu}+l < t_{\text{VLASS 2}}$.
Here, $l$ represents a modeled time lag which we increase between $-525$ and $+525\,$days in $75$ day increments in order to build our test samples.
We refer to these test samples of flaring sources as \textit{flaring (l day lag)} throughout the rest of this work, e.g., the flaring (0 day lag) sample incorporates no delay into the temporal association criteria, the flaring (-75 day lag) sample incorporates a negative lag of 75 days, and the flaring (+150 day lag) sample incorporates a delay of 150 days.
The test samples used in this work are summarized in Table \ref{tab:radsamp}.
The $\pm$ 525 day maximum shift applied to the temporal association window is comparable to half a VLASS epoch, and importantly still has full coverage in the version of the IceCat-1 catalog used in this work, which tracks events between MJDs of $55695$ and $60231$. 
Our VLASS variables have observation dates in the range $58004< \text{MJD} < 59631$, meaning any neutrino events we temporally associate with the VLASS sources must have been detected at $57479<\text{MJD}<60156$\footnote{Or, to put it another way, IceCat-1 covers events between May 2011 and October 2023, while our temporal association tests will only consider neutrinos detected between April 2016 and July 2023.}.


\renewcommand{\arraystretch}{1.25}
\begin{deluxetable}{lcc}
    \tabletypesize{\footnotesize}
    \tablecaption{Radio test samples used in this work
    \label{tab:radsamp}}
    \tablehead{
    \colhead{Sample Name} & \colhead{$N_{\text{radio}}$} & \colhead{$t_{\nu}$}\\
    \colhead{(1)} & \colhead{(2)} & \colhead{(3)}}
    \startdata
        Full variables  & $1,928$ & any\\
        Flaring ($l$ day lag)\tablenotemark{a} & $1,093$ & $t_{\text{VLASS 1}} < t_{\nu}+l < t_{\text{VLASS 2}}$
    \enddata
    \tablecomments{The name of the test sample (1); the number of radio sources in the sample (2), the neutrino event time, $t_{\nu}$, required in order to be associated with this sample.
    }
    \tablenotetext{a}{There are 15 flaring samples in total, having $l=$ -525, -450, -375, -300, -225, -150, -75, 0, +75, +150, +225, +300, +375, +450, or +525$\,$days.}
\end{deluxetable}

We match all of our samples of VLASS variables to the IceCat-1 data.
For the full variable sample using only a spatial association we find $446$ associations with IceCat-1, consisting of $381$ radio sources consistent with $138$ neutrino events.
When considering radio flares that are spatially and temporally associated with neutrinos, the flaring (0 day lag) sample has $64$ associations with IceCat-1, with $63$ distinct radio sources and $28$ neutrino events.
As we vary the lag time, the number of radio-neutrino associations increases with $l$ from \text{$N_{\text{assoc}}=44$} at $l=-525\,$days to $N_{\text{assoc}}=73$ at $l=+375\,$days before decreasing again to $N_{\text{assoc}}=52$ by $l=+525\,$days.
A similar trend is seen for the counts of radio sources, $N_{\text{radio}}$, and neutrinos, $N_{\nu}$, in that both broadly increase as the time lag is increased from $-525\,$days up to $+375\,$days before starting to decrease again.
We report the observed numbers for all of our samples in Table \ref{tab:results}.


\renewcommand{\arraystretch}{1.3}
\begin{table*}[]
    \centering
    \caption{Observed and expected counts for radio-neutrino associations, radio sources, and neutrino events.
    \label{tab:results}}
    \begin{tabular}{l|ccc|ccc|ccc}
        \hline\hline
        & \multicolumn{3}{c|}{radio-neutrino associations} & \multicolumn{3}{c|}{radio sources} & \multicolumn{3}{c}{neutrino events}\\
        \hline
        Sample Name & \hspace{1em}$N$\hspace{1em} & $\mu\pm\sigma$ & \hspace{0.5em}$p$-value\hspace{0.5em} & \hspace{1em}$N$\hspace{1em} & $\mu\pm\sigma$ & \hspace{0.5em}$p$-value\hspace{0.5em} & \hspace{1em}$N$\hspace{1em} & $\mu\pm\sigma$ & \hspace{0.5em}$p$-value\hspace{0.5em}\\
        \hline
        Full variables & $446$ & $445\pm19$ & $0.471$ & $381$ & $384\pm15$ & $0.572$ & $138$ & $147\pm7$ & $0.769$\\
        Flaring ($-525$ day lag) \hspace{2em} & $44$ & $47\pm7$ & $0.640$ & $42$ & $44\pm6$ & $0.660$ & $21$ & $24\pm3$ & $0.847$\\
        Flaring ($-450$ day lag) & $50$ & $47\pm7$ & $0.305$ & $48$ & $45\pm6$ & $0.305$ & $23$ & $23\pm3$ & $0.557$\\
        Flaring ($-375$ day lag) & $53$ & $47\pm7$ & $0.185$ & $52$ & $47\pm6$ & $0.223$ & $27$ & $25\pm3$ & $0.283$\\
        Flaring ($-300$ day lag) & $51$ & $48\pm6$ & $0.299$ & $51$ & $48\pm6$ & $0.318$ & $25$ & $25\pm3$ & $0.546$\\
        Flaring ($-225$ day lag) & $62$ & $53\pm7$ & $0.092$ & $61$ & $52\pm7$ & $0.094$ & $31$ & $26\pm3$ & $0.099$\\
        Flaring ($-150$ day lag) & $65$ & $53\pm7$ & $0.035$ & $64$ & $51\pm7$ & $0.037$ & $30$ & $26\pm3$ & $0.158$\\
        Flaring ($-75$ day lag) & $64$ & $53\pm7$ & $0.046$ & $63$ & $51\pm7$ & $0.054$ & $29$ & $26\pm3$ & $0.183$\\
        Flaring ($0$ day lag) & $64$ & $52\pm7$ & $0.034$ & $63$ & $51\pm6$ & $0.031$ & $28$ & $26\pm3$ & $0.260$\\
        Flaring ($+75$ day lag) & $63$ & $53\pm7$ & $0.053$ & $63$ & $51\pm6$ & $0.039$ & $27$ & $26\pm3$ & $0.413$\\
        Flaring ($+150$ day lag) & $65$ & $50\pm7$ & $0.014$ & $64$ & $49\pm7$ & $0.015$ & $30$ & $26\pm3$ & $0.157$\\
        Flaring ($+225$ day lag) & $64$ & $50\pm7$ & $0.019$ & $63$ & $49\pm6$ & $0.021$ & $29$ & $26\pm3$ & $0.204$\\
        Flaring ($+300$ day lag) & $67$ & $53\pm7$ & $0.019$ & $65$ & $52\pm7$ & $0.033$ & $31$ & $27\pm3$ & $0.148$\\
        Flaring ($+375$ day lag) & $73$ & $58\pm7$ & $0.016$ & $71$ & $57\pm7$ & $0.026$ & $31$ & $27\pm3$ & $0.173$\\
        Flaring ($+450$ day lag) & $60$ & $55\pm7$ & $0.199$ & $58$ & $54\pm7$ & $0.264$ & $27$ & $27\pm3$ & $0.529$\\
        Flaring ($+525$ day lag) & $52$ & $54\pm7$ & $0.542$ & $51$ & $53\pm7$ & $0.608$ & $22$ & $26\pm4$ & $0.873$\\
        \hline
        \multicolumn{10}{l}{\vspace{-1em}}\\ 
        \multicolumn{10}{l}{\hspace{1em} \textsc{Note.}\ \textemdash\, For each test sample we list: the observed count ($N$), the mean ($\mu$) and standard deviation ($\sigma$) of the counts}\\
        \multicolumn{10}{l}{from the Monte Carlo simulations, and the $p$-value for observing $N$ based on the Monte Carlo distribution, for the number}\\
        \multicolumn{10}{l}{of radio-neutrino associations, as well as the number of radio sources and neutrino events in those associations.}
    \end{tabular}
\end{table*}

As might be expected given the relatively long cadence of VLASS, there is substantial overlap between our samples of flaring sources and their neutrino associations.
There are $99$ unique radio sources across all of our flaring samples, and $46$ distinct neutrino events with which those sources are associated.
Of these, $6$ radio sources are in as few as two flaring test samples, with the majority being in 10 or more samples.
Similarly, $1$ neutrino event is found in only two of the flaring test sets, and nearly half of neutrinos appear in at least 10 samples.
It is also important to note that using our definitions of associations can result in multiple candidate radio counterparts for a single neutrino event (e.g., see Figure \ref{fig:icecube-x-vlass}c).
This is to be expected when considering the large uncertainties in the neutrino position, and the on-sky density of sources in VLASS.
For instance, the median radius of the $90\,\%$ confidence positional uncertainty in the IceCat-1 data is $\sim80^{\prime}$, corresponding to a typical on-sky footprint of $\sim5.6\,\text{deg}^{2}$ in which the neutrino's source origin is expected to lie.
VLASS has a source density of $\rho\sim53\,\text{deg}^{-2}$, so a typical neutrino detection is expected to have $\sim300$ candidate counterparts in VLASS.
Radio sources that vary by $>50\,\%$ in flux density are far rarer, with our data here having a typical on-sky source density of $\rho\sim0.06\,\text{deg}^{-2}$, corresponding to an average of $0.3$ of these objects expected to be associated with any given IceCat-1 detection.
For those neutrinos where there are multiple associated radio sources, only one of these, if any, can be the true point of origin of the neutrino.
It is also possible for some radio sources to be associated with multiple neutrino events\textemdash especially when there is no time constraint required for the association.
Such sources are of interest on their own as multiple neutrino associations with an object reduce the likelihood of the associations being coincidental.
We will study these candidate multi-neutrino radio variables in depth in a future work.
Our aim in this paper is not to identify the correct source of origin of neutrino events, but rather to look for potential trends in whether variable radio sources as a population might be consistent with neutrinos more often than would be expected if those associations were due to random chance.


\subsection{Comparisons with the Expected Random Background Contribution} \label{ssec:stats}
\subsubsection{Numbers of Associations} \label{ssec:assoc-counts}

In order to test whether the observed numbers of associations between variable radio sources and HE neutrinos are significant, we perform a $1,000$ iteration Monte Carlo (MC) simulation for each test sample.
In each iteration, $N$ random sky coordinates within the VLASS footprint are used to simulate the radio population, where $N$ is the number of radio sources in the real test sample (i.e., $N=1,928$ for the full variable sample and $N=1,093$ for the flaring samples).
In practice these simulated sky positions are achieved by randomizing the list of R.A. coordinates for the radio source test sample, an approach that ensures the simulated radio sources have the same declination distribution as the real radio sources.
As the VLASS footprint is simply defined as $\delta > -40^{\circ}$, retaining the declination distribution of the real radio samples will result in simulated positions that are also within the VLASS footprint.
Moreover, maintaining the radio source declination distribution mitigates the potential for the declination bias of the IceCat-1 events (the clustering of red ellipses toward the equator in Figure \ref{fig:aitoff-icecube-vlass}) to impact our comparisons of real and simulated radio-neutrino associations.
Shuffling the R.A. of the radio sources also has the secondary effect of randomizing the time-domain information for the samples where we are also searching for evidence of a temporal correlation.
Any real temporal correlation with the neutrinos that may exist will be lost in the mock data as the sky positions will be random, allowing the Monte Carlo to effectively simulate a random background of flaring sources for the flaring and lagged-flaring samples.

\begin{figure}
    \centering
    \includegraphics[width=\columnwidth]{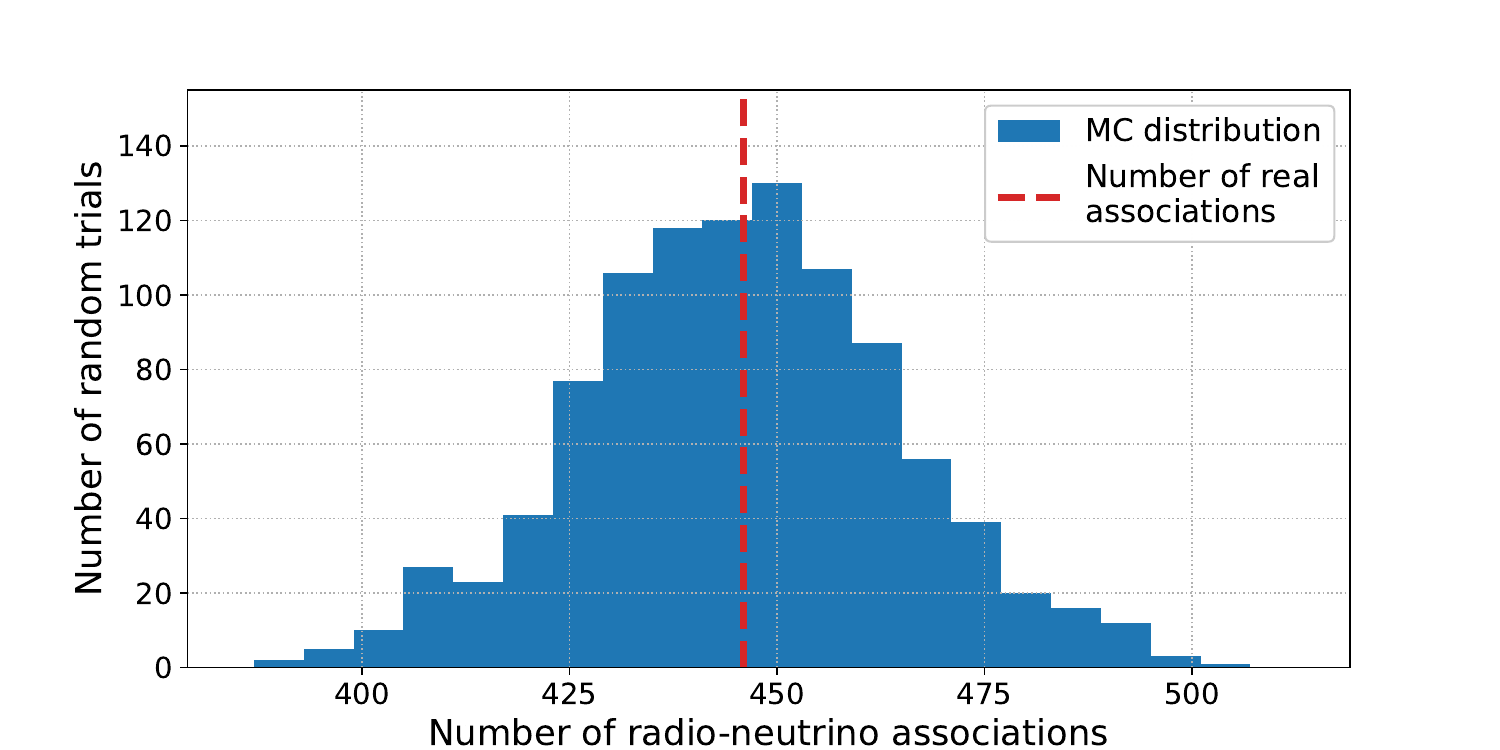}
    \caption{The number of associations between VLASS radio sources and IceCube neutrinos for the 'full variables' sample where we look for positional associations only with IceCat-1 neutrino events, not temporal.
    The blue histogram shows the distribution of the association counts in the 1,000 run Monte Carlo simulation, and the red dashed line shows the observed number of associations.}
    \label{fig:1dmc}
\end{figure}

A real correlation between radio sources and neutrinos should show an excess number of associations relative to random background, and the significance of any excess can be estimated as a $p$-value defined by the fraction of MC iterations with more associations than the observed value.
In Figure \ref{fig:1dmc} we show the distribution of radio-neutrino association counts obtained by each iteration of the MC simulation for our full variables sample (which imposes no time constraint on the neutrino detection) as a blue histogram.
The red dashed vertical line on Figure \ref{fig:1dmc} shows the real number of associations observed for the full variables sample.
The MC distribution for the full variables sample has a mean ($\mu$) and standard deviation ($\sigma$) of $445\pm19$, consistent with the observed value of $446$ associations.
Of the $1,000$ MC iterations, $471$ resulted in $\geq446$ associations, suggesting a $p$-value of $0.471$ for the observed number of associations in the full variable sample.
We therefore see no evidence of a purely spatial correlation between the full variables sample and HE neutrinos in IceCat-1.

\begin{figure}
    \centering
    \includegraphics[ width=\columnwidth]{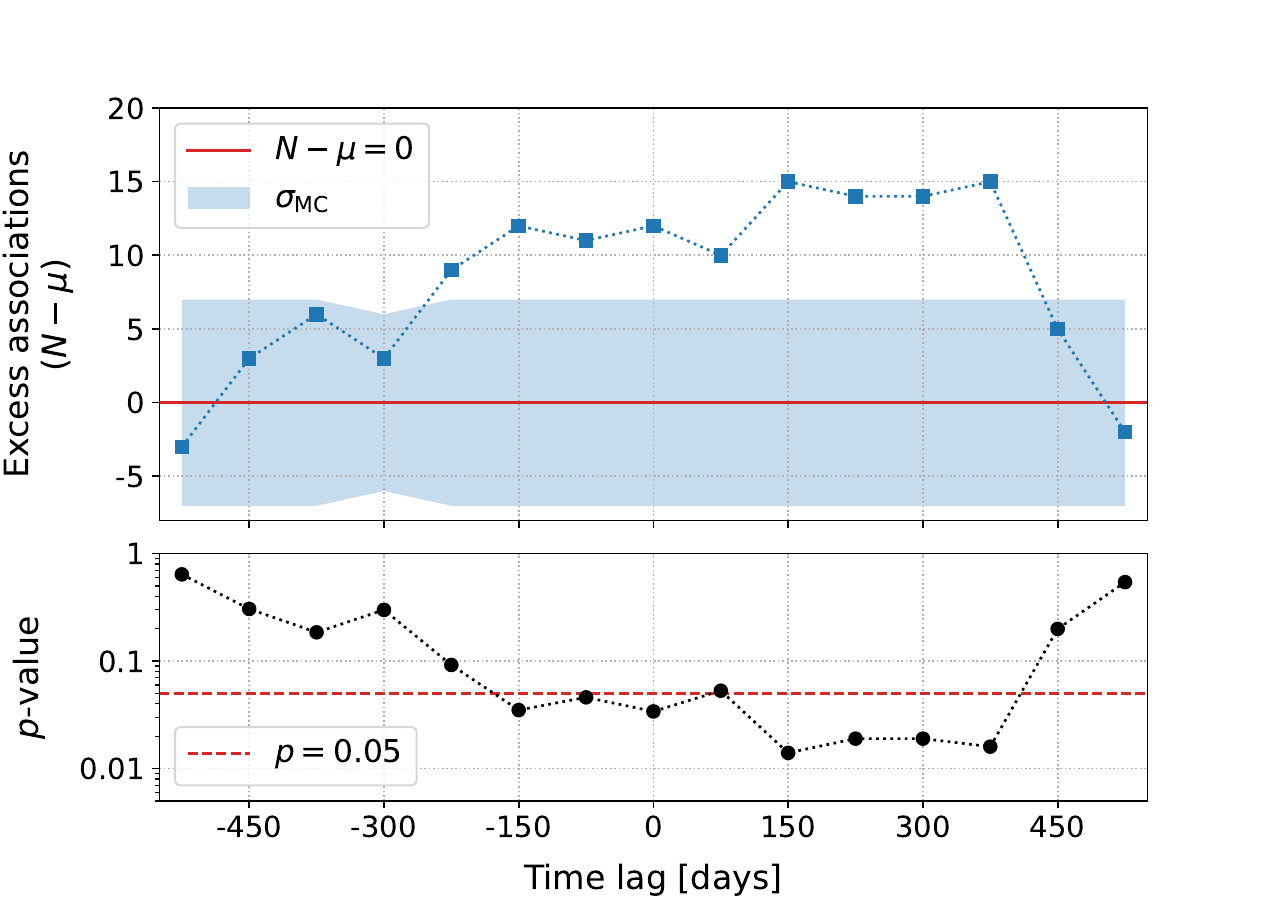}
    \caption{The excess numbers of radio-neutrino associations for flaring radio sources as a function of the time lag used for the temporal correlation.
    The upper panel shows the observed number of associations relative to the mean value from the Monte Carlo simulation, $N_{\text{assoc}}-\mu_{\text{assoc, MC}}$.
    The lower panel shows the $p$-value of the observed excess obtained from the Monte Carlo distribution.
    The shaded blue region in the upper panel shows the standard deviation from the Monte Carlo distributions, and the solid red horizontal line shows zero excess.
    The horizontal dashed red line in the lower panel highlights $p=0.05$.}
    \label{fig:association-excess}
\end{figure}

When considering our flaring samples we find measurable excess numbers of radio-neutrino associations when the time lag is in the range $-150\,\text{days} \leq l \leq +375\,\text{days}$.
In the upper panel of Figure \ref{fig:association-excess} we plot the excess number of associations, defined as $N-\mu$, where $N$ is the real observed count, versus the time lag.
The shaded blue region highlights the standard deviation of the MC simulation for each sample.
In the lower panel of Figure \ref{fig:association-excess} we plot the $p$-values based on the MC distributions.
For $-150\,\text{days} \leq l \leq +75\,\text{days} $ we see $\sim12$ excess associations over the expected random background counts with $p\lesssim 0.05$.
At lags in the range $+150\,\text{days} \leq l \leq +375\,\text{days}$ we see $\sim 15$ additional associations over the expected number with $p<0.02$.
When the time lag is $<-150\,\text{days}$ or $>+375\,\text{days}$ the observed real number of associations is found to be consistent with the expected random background counts.
The means, standard deviations, and $p$-values for all of our test samples are listed in Table \ref{tab:results} alongside the real observed counts.

Testing multiple time lags does increase the chances of seeing an excess number of associations due to random chance.
In a basic multi-trial experiment one could account for this effect by adjusting the calculated $p$-values by a Bonferroni correction factor.
However, the long cadence of VLASS  observations ($\sim2.5\,$ years) means that each lag tested is not independent from the others\textemdash radio-neutrino associations are frequently in multiple test samples.
While this likely acts to lower the impact of the Bonferroni correction needed (the 15 different lags are not 15 independent trials), that non-independence of the samples also makes calculating the correction factor challenging.
While not definitive, our results show a coherent picture where a spatiotemporal correlation exists at $\sim2\sigma$ confidence between flaring radio sources in VLASS and HE neutrinos in IceCat-1 when accounting for a time lag of $-150\,\text{days} \lesssim l \lesssim +375\,\text{days}$ between the neutrino detection and the radio flare.
The strongest signal is seen at lags of $+150\,\text{days} \leq l \leq +375\,\text{days}$, suggesting there may be a preferred lag between neutrino and radio emission on similar timescales to that seen between $\gamma$-ray and radio emission in blazars \citep[e.g.,][]{MaxMoerbeck2014, Fuhrmann2014}.
A catalog of all the radio-neutrino associations for our flaring samples is made available to the wider scientific community as described in Appendix \ref{ap:assocdata}.

\subsubsection{Angular Separations Between Associations} \label{ssec:angseps}

\begin{figure}
    \centering
    \includegraphics[trim={0 0 0 1cm}, clip, width=\columnwidth]{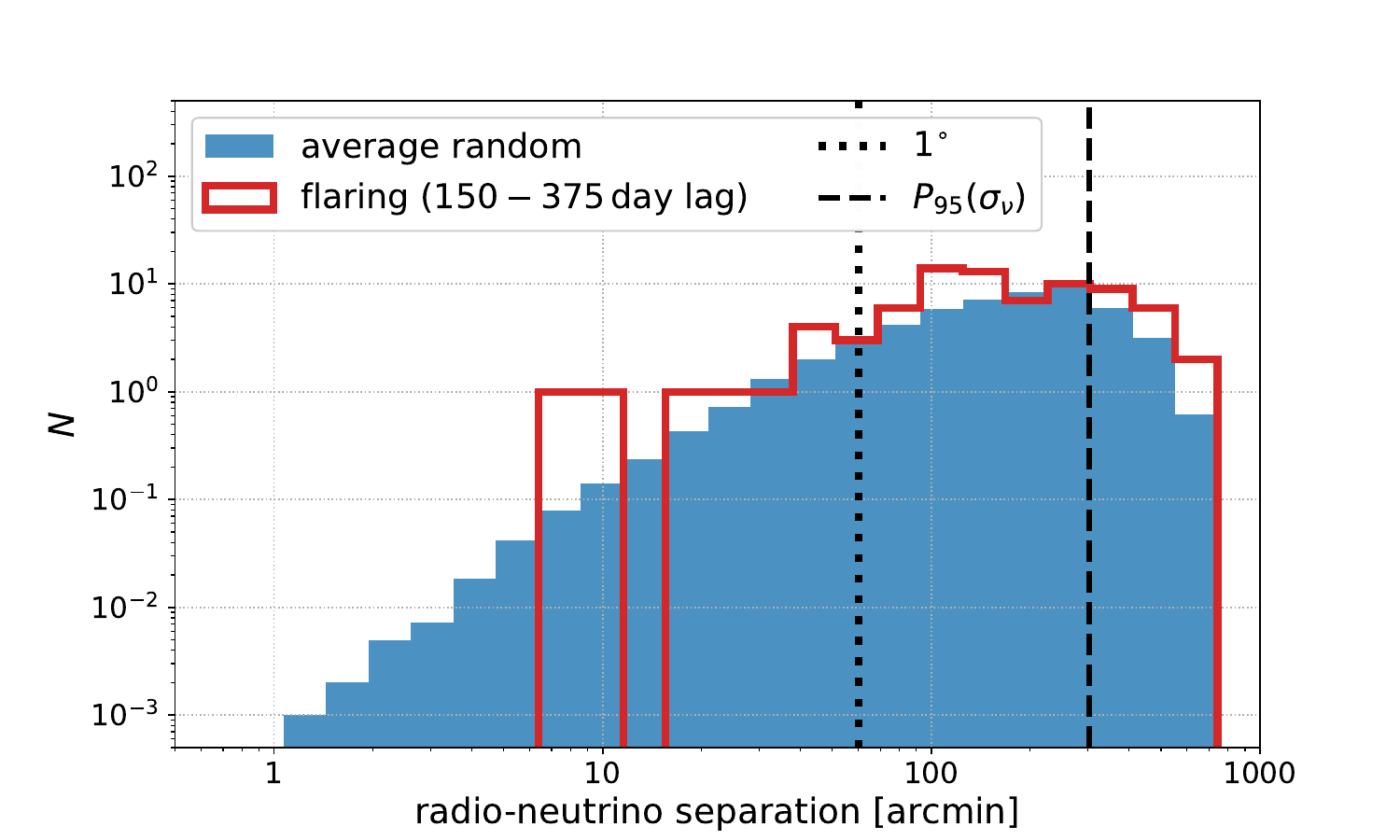}
    \caption{Distributions of angular separations between associated radio sources and neutrino events.
    The solid blue histogram shows random associations from Monte Carlo simulations. 
    Angular separations for real associations from the flaring samples with time lags of $+150$ to $+375\,$days are shown by the red solid line.
    The dashed black vertical line shows the 95th percentile of the positional errors exhibited by IceCat-1 events, $\sigma_{\nu}$, and the dotted black vertical line represents an angle of $1^{\circ}$.
    }
    \label{fig:angseps}
\end{figure}

If there is truly a correlation between variable radio sources and HE neutrinos then one might expect the on-sky angular separations between the radio and neutrino detections to be smaller than would be expected from associations that were purely the result of the random background distributions.
In Figure \ref{fig:angseps} we compare the real observed distributions of angular separations between the radio source and the neutrino detection with the expected random background distribution of angular separations.
The red solid line in Figure \ref{fig:angseps} shows the $79$ radio-neutrino associations that appear in any of our flaring samples with time lags in the range $+150\,\text{days} \leq l \leq +375\,\text{days}$, i.e., those where we see an excess number of associations at $p<0.02$.
Each association is only included once in Figure \ref{fig:angseps}.
The blue histogram in Figure \ref{fig:angseps} shows the distributions from the Monte Carlo simulations from all four of those flaring samples with $+150\,\text{days} \leq l \leq +375\,\text{days}$ weighted by a factor of $1/4000$, and should thus represent the expected number of random associations with a given angular separation.
The black dotted vertical line represents an angular separation of $1^{\circ}$ while the black dashed vertical line shows the $95$th percentile of positional uncertainties in IceCat-1, $\sigma_{\nu}$.
The median angular separation of real observed radio-neutrino associations in these flaring samples is $\sim140^{\prime}$ compared to $\sim180^{\prime}$ for the random associations.

\begin{figure}
    \centering
    \subfigure[]{\includegraphics[trim={0 0 0 1cm}, clip, width=\columnwidth]{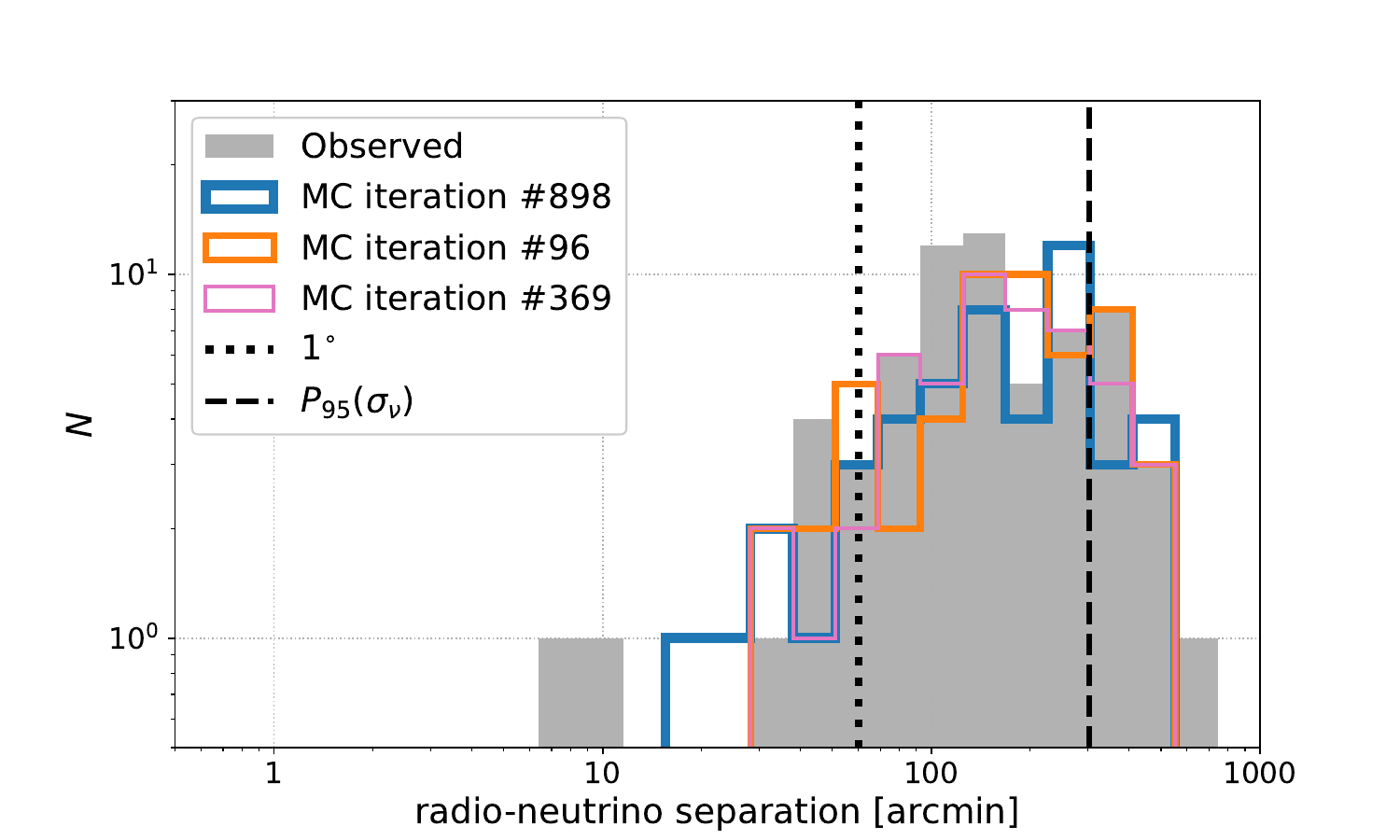}}
    \subfigure[]{\includegraphics[trim={0 0 0 1.5cm}, clip, width=\columnwidth]{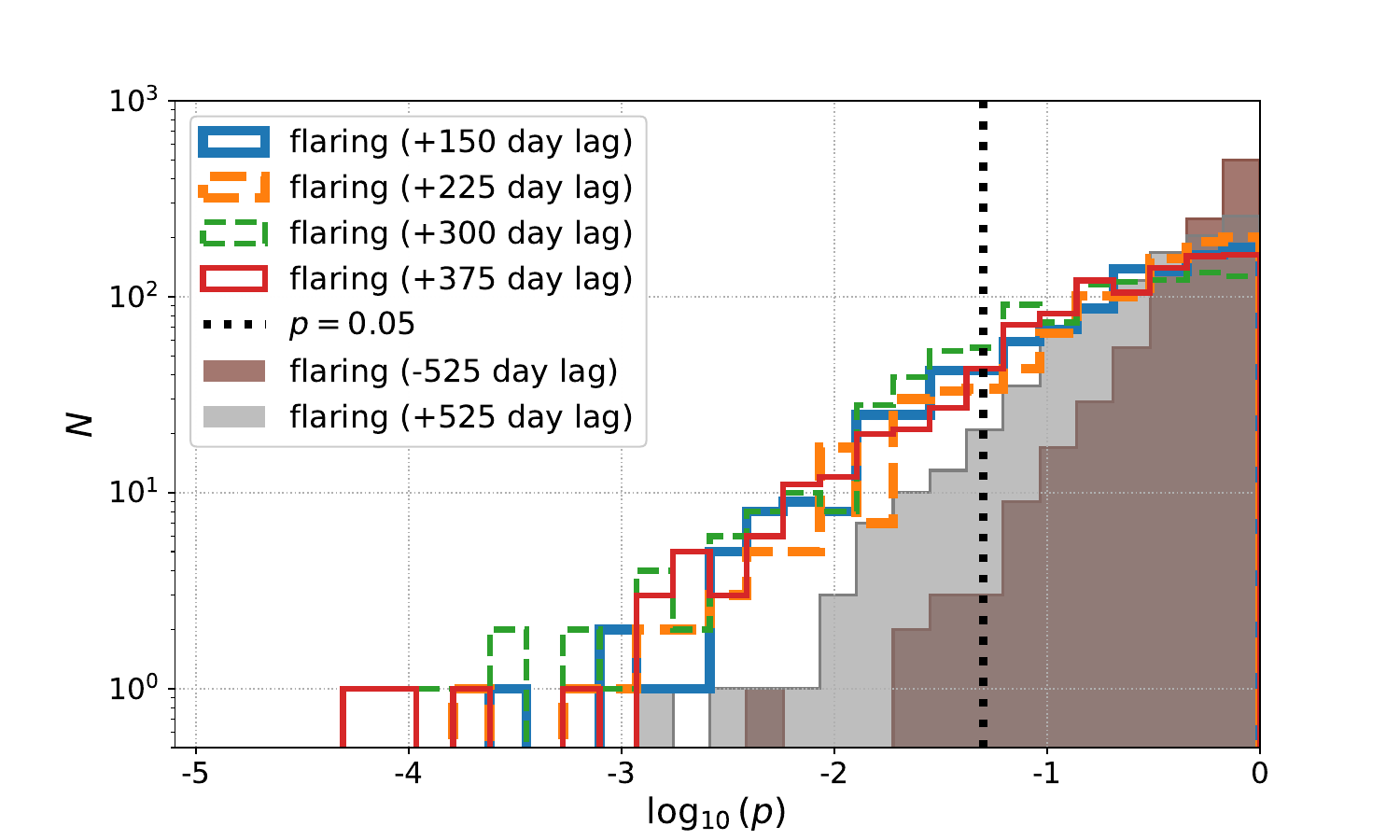}}
    \caption{Panel (a) shows the distribution of real observed radio-neutrino angular separations (gray solid histogram) for our flaring (+150 day lag) sample alongside the distributions from three example iterations from the Monte Carlo (blue, orange, and pink lines).
    Panel (b) shows the distribution of $p$-values obtained from two-sample KS tests comparing the observed radio-neutrino angular separations with the simulated data for each of the 1,000 MC iterations. 
    The distributions of $p$-values are shown for the flaring samples with $+150$ (blue solid line), $+225$ (orange dashed line), $+300$ (green dashed line), and $+375$ (red solid line) day lags.
    The brown and gray solid histograms show the distributions for the flaring samples with $-525\,$day and $+525\,$day lags, respectively, samples for which no excess of radio-neutrino associations are seen.
    }
    \label{fig:angsepstats}
\end{figure}

The smaller median separation of real associations between radio flares and neutrinos compared to those seen in the Monte Carlo simulations is likely driven by the couple of real associations with $\sim10^{\prime}$ angular offsets.
From Figure \ref{fig:angseps} we might expect an average value of $\sim 0.1$ associations at such separations due to random chance given the number of radio sources in the flaring sample.
It is of course impossible to have non-integer association counts, so we further compare the distributions of real observed angular separations to the random distribution from each iteration of the respective Monte Carlo simulation.
For example, in Figure \ref{fig:angsepstats}a we show the distribution of the radio-neutrino separations for the flaring ($+150$ day lag) as a gray histogram alongside the distributions obtained from three random iterations of the MC simulation (colored lines) for that test sample.
For the sake of clarity we only show three iterations for one test sample as an example here, but there are naturally $1,000$ such comparisons to make for each of our test samples.

\renewcommand{\arraystretch}{1.25}
\begin{deluxetable}{lcc}
    \tabletypesize{\footnotesize}
    \tablecaption{Radio-neutrino angular offset statistics
    \label{tab:angseps}}
    \tablehead{
    \colhead{Sample} & \colhead{$\psi_{\text{med}}$} & \colhead{$f_{p<0.05}$}\\
    \colhead{(1)} & \colhead{(2)} & \colhead{(3)}}
    \startdata
        Flaring ($-525$ day lag) & $154^{\prime}$ & $0.008$\\
        Flaring ($+150$ day lag) & $140^{\prime}$ & $0.136$\\
        Flaring ($+225$ day lag) & $139^{\prime}$ & $0.124$\\
        Flaring ($+300$ day lag) & $139^{\prime}$ & $0.188$\\
        Flaring ($+375$ day lag) \hspace{5em} & \hspace{1em}$152^{\prime}$\hspace{1em} & $0.136$\\
        Flaring ($+525$ day lag) & $163^{\prime}$ & $0.051$\vspace{0.3em}
    \enddata
    \tablecomments{
    The name of the test sample (1); the median angular offset between radio sources and associated neutrino events (2);  and the fraction of Monte Carlo simulations where the KS test comparing the radio-neutrino angular separations for the real and random samples returns $p<0.05$ (3).
    }
\end{deluxetable}

As a point of comparison, we also perform this analysis for the two flaring samples with time lags of $\pm525\,$days.
These two samples show no excess in the number of radio-neutrino associations compared to the expected number of random associations, and bracket the time lags where evidence of such an excess is seen (see Section \ref{ssec:assoc-counts}, especially Figure \ref{fig:association-excess}, and Table \ref{tab:results}).
In Figure \ref{fig:angsepstats}b, the flaring ($-525\,$day lag) sample is shown as a brown solid histogram and the flaring ($+525\,$day lag) sample is shown as a solid gray histogram.
For our flaring samples with $+150\,\text{days} \leq l \leq +375\,\text{days}$, the fraction of KS tests that return $p<0.05$, $f_{p<0.05}$,
is higher ($0.12<f_{p<0.05}<0.19$) than in the flaring samples with lags of $-525\,$days ($f_{p<0.05}=0.01$) and $+525\,$days ($f_{p<0.05}=0.05$).
The $f_{p<0.05}$ statistics are summarized in Table \ref{tab:angseps} alongside the median radio neutrino separations for each of these samples.
While not conclusive, our investigation here suggests that when accounting for a potential time lag between the neutrino and radio emission, real radio flares are typically closer on the sky to the position of the neutrinos with which they are potentially associated than might be expected due to random chance.
We will investigate this further by analyzing the two-point correlation function between radio flares and HE neutrinos in a future work where we have additional observational data.


\begin{figure*}
    \centering
    \subfigure[Flaring ($+150$ day lag)]{\includegraphics[width=\columnwidth]{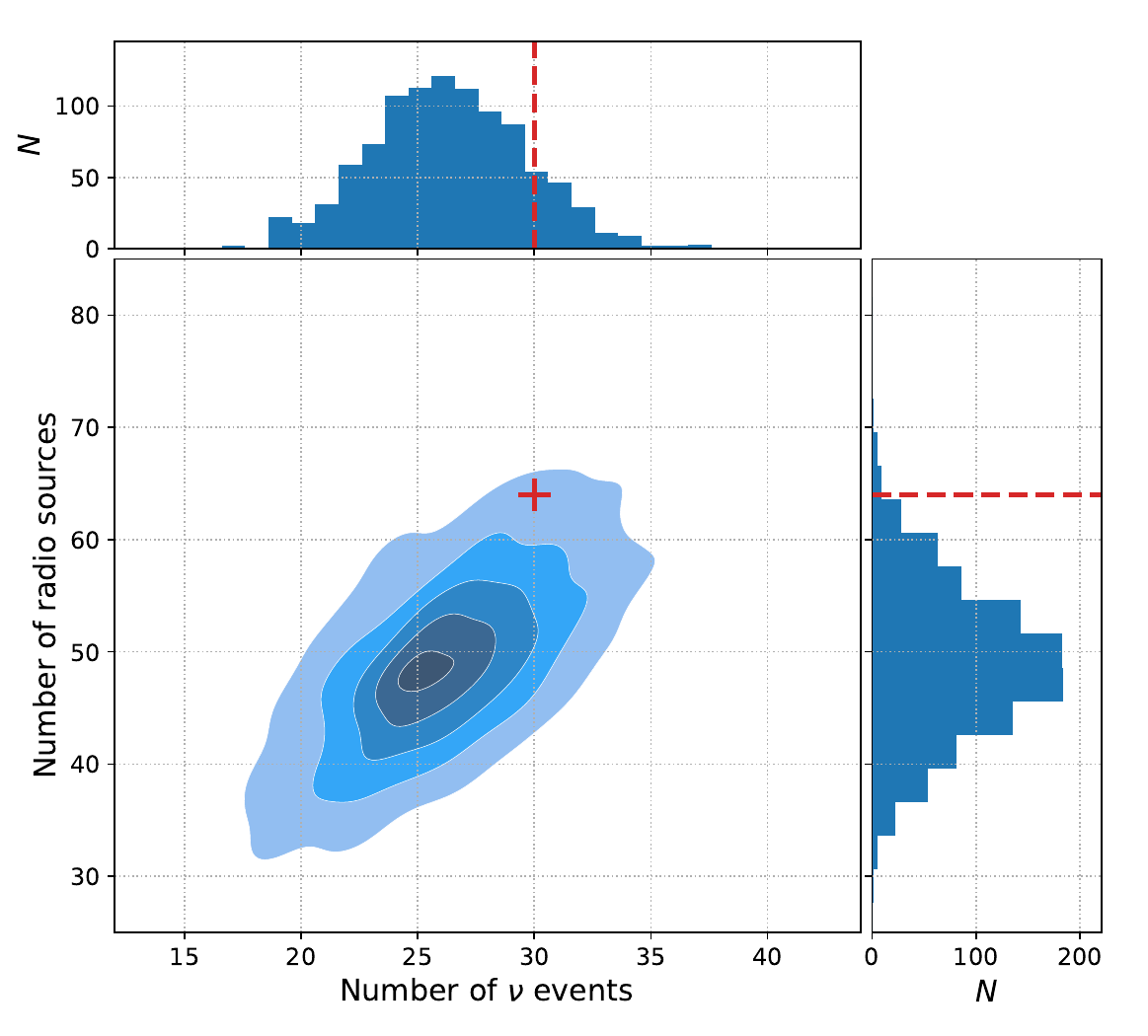}}
    \subfigure[Flaring ($+225$ day lag)]{\includegraphics[width=\columnwidth]{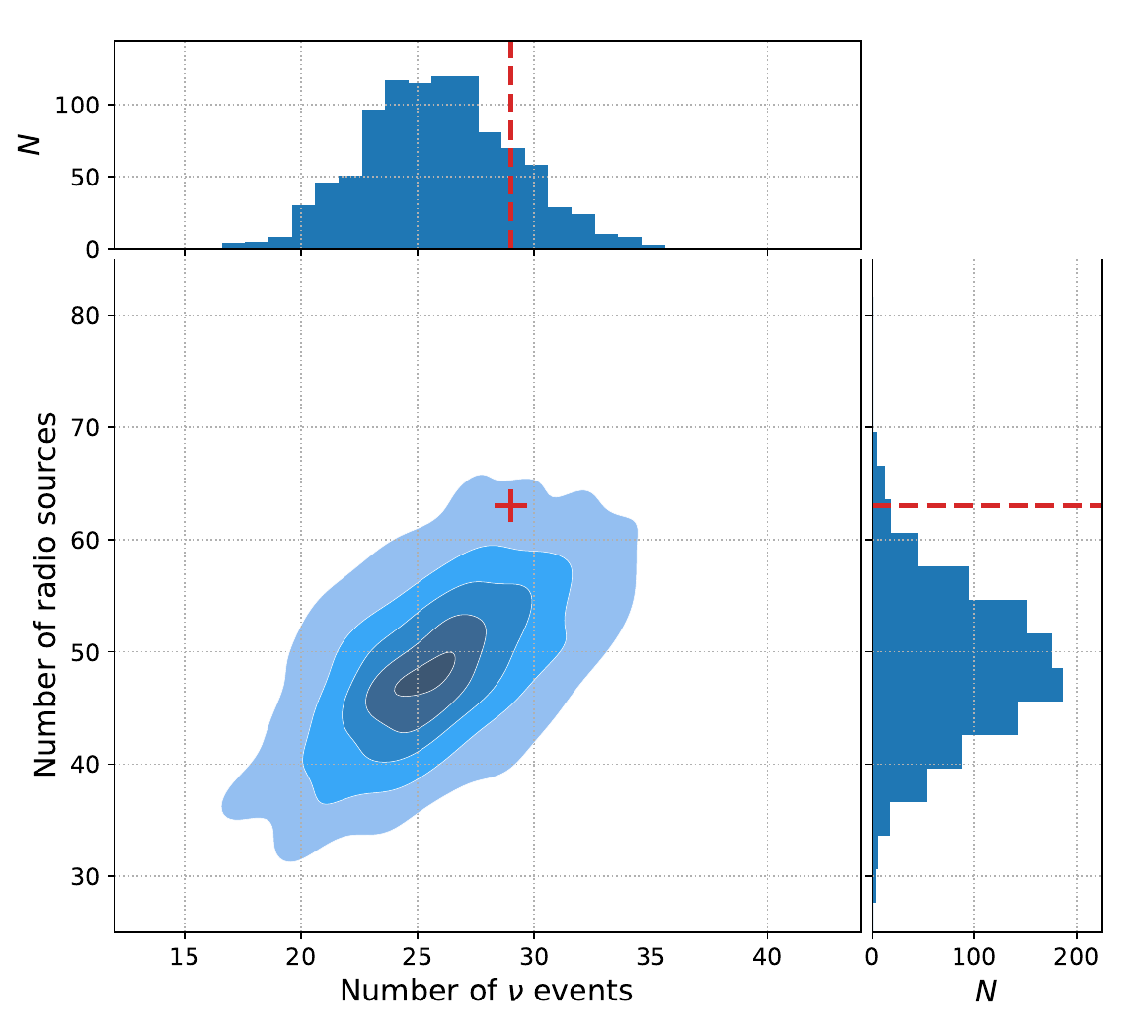}}
    \subfigure[Flaring ($+300$ day lag)]{\includegraphics[width=\columnwidth]{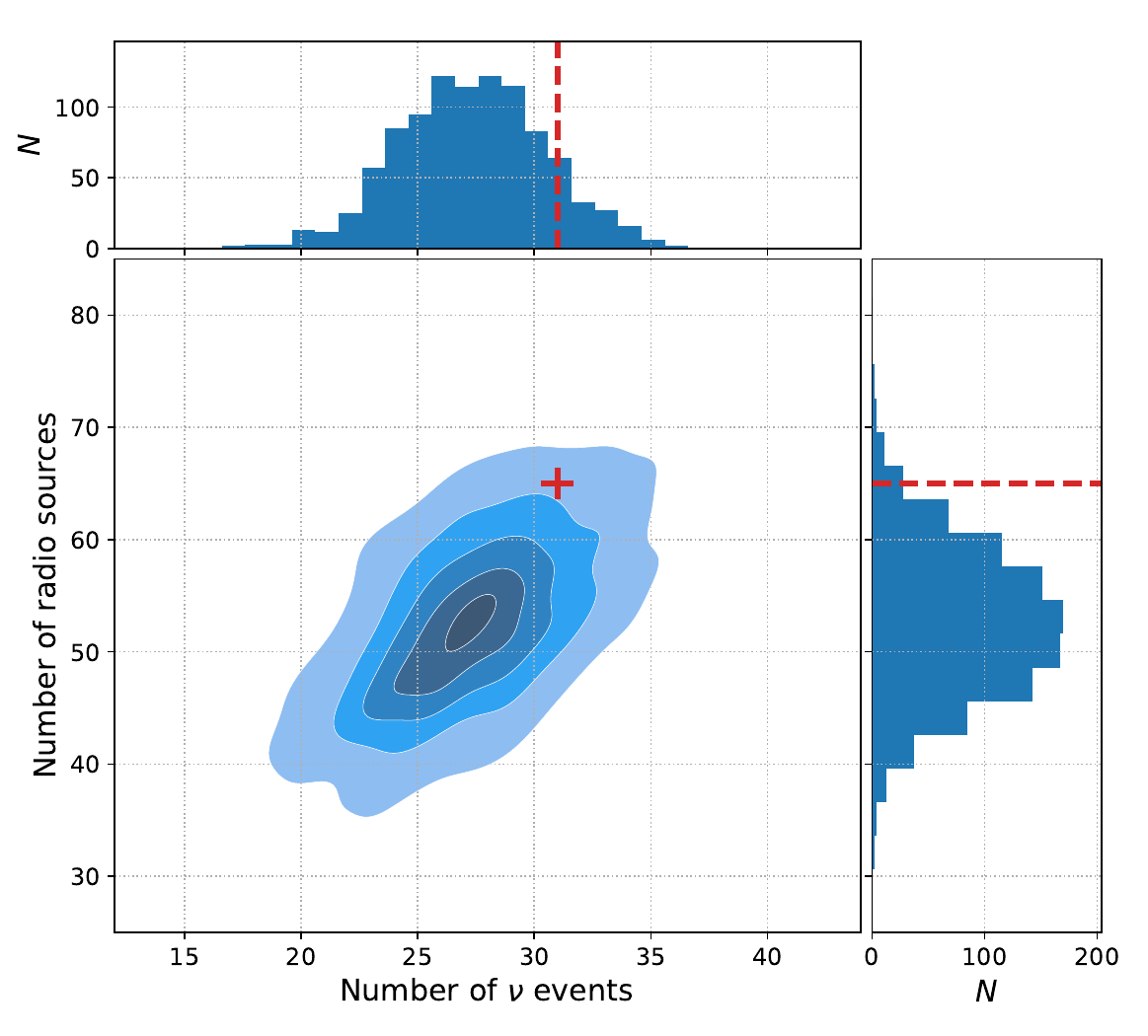}}
    \subfigure[Flaring ($+375$ day lag)]{\includegraphics[width=\columnwidth]{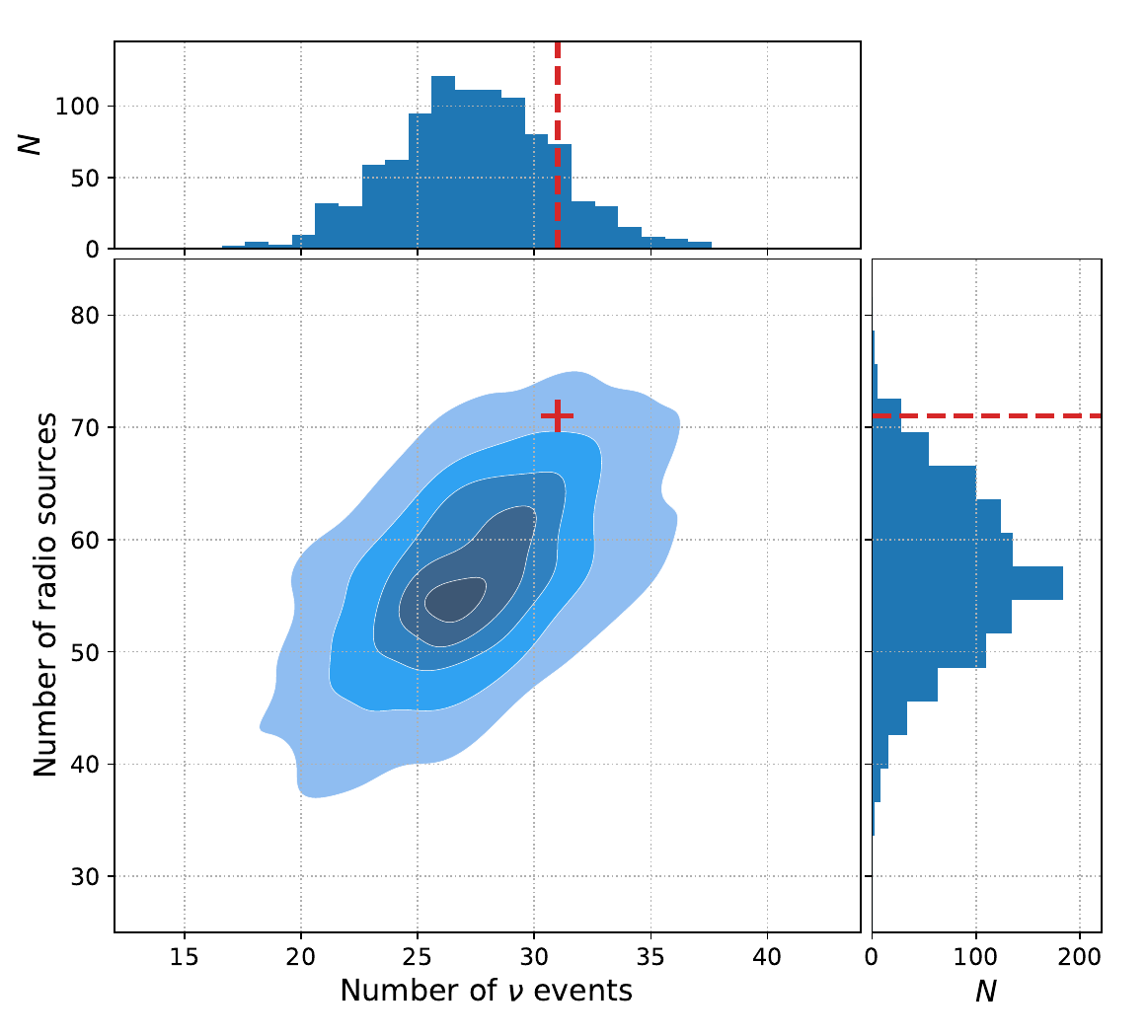}}
    \caption{The expected background distributions of radio source  and neutrino event counts for the four test samples with a higher than expected number of radio-neutrino associations with $p<0.02$. 
    The panels show the samples of flares with lags between the neutrino detection and the VLASS time window of $+150$ days (a), $+225$ days (b), $+300$ days (c), and $+375$ days (d).
    The contours in each panel show the $95$th, $75$th, $50$th, $25$th, and $5$th percentiles from the PDF estimated by the Monte Carlo simulation.
    The red cross and dashed lines show the position of the observed numbers of neutrino events and associated radio sources.}
    \label{fig:2dmc}
\end{figure*}

\subsubsection{Radio and Neutrino Counts} \label{ssec:radio-neutrino-counts}

In Figure \ref{fig:2dmc}, we split the associations into the number of neutrino events and number of flaring radio sources and plot the individual distributions of these counts for the four test samples where we see the strongest evidence for an excess number of radio-neutrino associations, the flaring samples with time lags of $l=+150, +225, +300,$ and $+375\,$days.
In all four tests, the $p$-values for the excess number of radio sources are $<0.05$,
while the excess number of neutrino events have $0.15 \lesssim p \lesssim0.2$.
The observed and expected radio and neutrino count statistics for all sixteen of our test samples are summarized in Table \ref{tab:results}.
The significance of the excess number of associations between radio flares and neutrinos is clearly driven by the number of radio sources rather than the neutrino counts.
This is likely due to the rarity of neutrino detections relative to the density of flaring radio sources on the sky.

\section{Discussion} \label{sec:discussion}


\subsection{Multiwavelength Properties of Radio Sources Consistent with Neutrino Events} \label{ssec:sourceproperties}



Additional information about flaring radio sources can be gleaned by studying their multiwavelength properties.
In \citet{Gordon2025} we obtained for the VLASS variables:
\begin{itemize}
    \item redshift measurements from the Sloan Digital Sky Survey data release 16 \citep[SDSS DR16,][]{York2000, Ahumada2020} and the DESI Legacy Surveys data release 8 \citep[LS DR8,][]{Dey2019, Duncan2022};
    \item infrared counterparts from the Wide-field Infrared Survey Explorer telescope's \citep[WISE,][]{Wright2010} AllWISE catalog \citep{Cutri2014};
    \item and $\gamma$-ray counterparts from the fourth Fermi Large Area Telescope data release 4 \citep[4FGL,][]{Abdollahi2020, Ballet2023}.
\end{itemize}
Throughout Section \ref{ssec:sourceproperties}, we use these data alongside the VLASS measurements to explore the multiwavelength properties of flaring radio sources potentially associated with HE neutrino events.
Explicitly we focus on the 77 distinct radio sources that are found in any flaring samples with time lags between $+150\,\text{days}$ and $+375\,\text{days}$, as these samples show the strongest signal for an excess of radio-neutrino associations in our analysis ($p<0.02$).

\subsubsection{Radio Brightness} \label{sssec:radflux}





\begin{figure}
    \centering
    \includegraphics[width=\columnwidth]{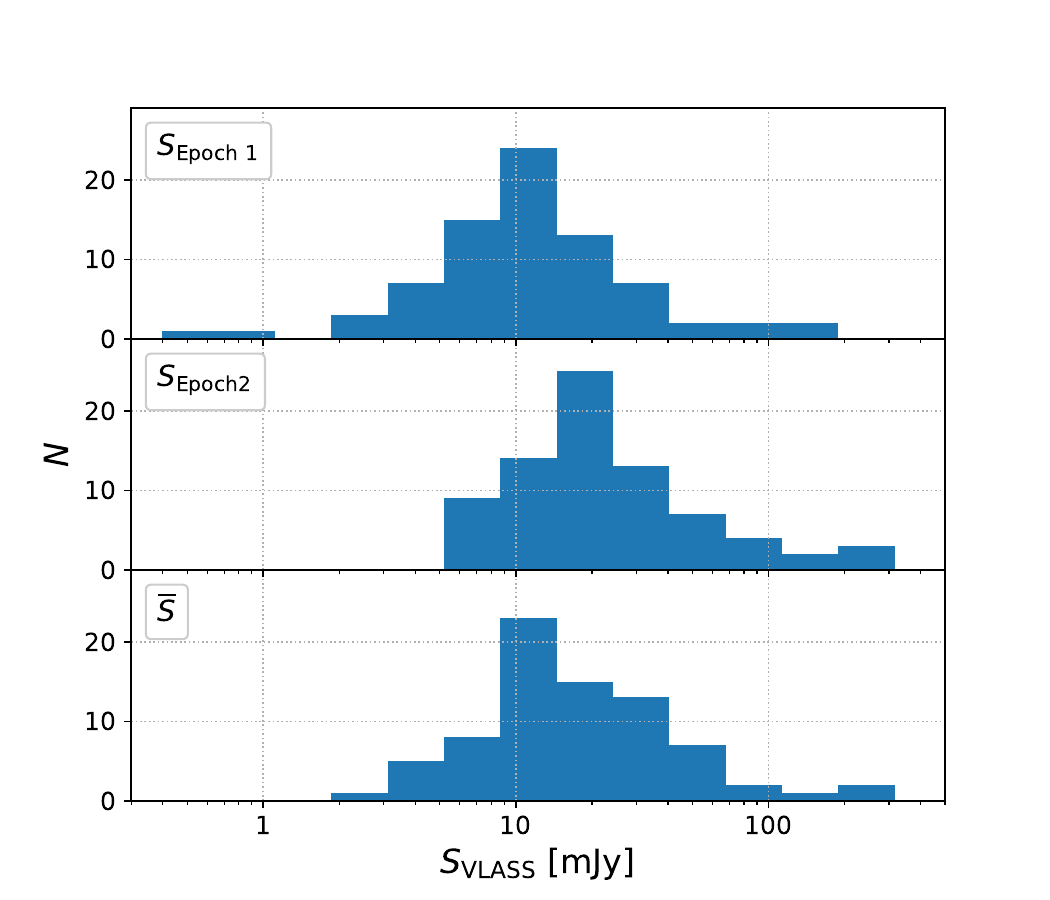}
    \caption{The VLASS flux densities, $S_{\text{VLASS}}$, for the 77 radio sources that appear in the flaring samples with time lags between $+150$ and $+375$ days.
    The upper panel shows the distribution of flux densities in Epoch 1, the middle panel the flux densities in Epoch 2, and the lower panel shows the distribution of the mean brightness from the two epochs of VLASS observations.}
    \label{fig:radiofluxes}
\end{figure}

One of the advantages of VLASS for the type of cross correlation analyses presented in this paper is that it provides variability information on much fainter objects than have historically been the target of radio time-domain surveys.
For instance, \citet{Hovatta2021} used observations from the Owens Valley Radio Observatory (OVRO) blazar monitoring program \citep{Richards2011} in their search for a correlation between radio flares and neutrinos.
Those blazars are selected from the Candidate Gamma Ray Blazar Survey \citep[CGRaBS,][]{Healy2008}, and have flat radio spectra with flux densities brighter than $65\,$mJy at $4.8\,$GHz.
The faintest detections in the VLASS variables catalog have brightnesses of $\sim 0.5\,$mJy, finding radio-variable sources that would not have been identified previously.
In Figure \ref{fig:radiofluxes}, we show the distribution of VLASS fluxes in Epoch 1 ($S_{\text{Epoch 1}}$; upper panel) and Epoch 2 ($S_{\text{Epoch 2}}$; middle panel), and the mean brightness across the two epochs ($\overline{S}$; lower panel) for the 77 radio sources in the flaring ($+150$ to $+375$ day lag) samples.
The vast majority of these objects have flux densities between a few and a few tens of mJy at $3\,$GHz.
Only four are brighter than $50\,$mJy, and none of these appear in the CGRaBS catalog.

It is possible that some very bright flaring sources may be inadvertently excluded from this analysis. 
While VLASS is an effective probe of the faint radio variable sky, the current VLASS imaging is known to have issues with the brightest sources in the survey.
For instance, image artifacts around objects brighter than a few hundred mJy can cause problems with source fitting \citep{Gordon2021}.
To account for such issues, the selection of VLASS variables in \citet{Gordon2025} was limited to objects that passed a visual inspection to make sure that  artifacts were not dominating the image.
One approach to navigating the potentially missing radio flares would be to more broadly incorporate radio time-domain information from multiple observing programs.
Such an approach would help maximize the number of independent radio-neutrino temporal associations counted, reducing the shot noise in the analysis.
However, this would be non-trivial compared to analyzing the relatively homogeneous data from a single survey.
For example, one would need to account for the different cadences of the radio observations, as well as the (relatively poorly constrained) potential time lag between the neutrino event and the radio flare.
That scale of work is beyond the scope of this paper\textemdash which presents a first attempt at studying radio-neutrino correlations with deep near-all-sky radio time-domain surveys\textemdash but provides an intriguing opportunity for further research in the near future.


\subsubsection{High Energy Electromagnetic Counterparts} \label{sssec:gammax}

Of particular interest here, the mechanisms for neutrino production are also typically expected to produce corresponding high energy photons.
We thus might expect flaring radio sources associated with neutrino production to also have a $\gamma$-ray counterpart.
Of the 77 flaring radio sources, we find Fermi counterparts to $3$ objects, or $4\,$\% of the sample.
While $\gamma$-rays are the most obvious byproduct of neutrino production, their extremely short wavelengths (a $1\,$GeV photon has a wavelength of $\sim10^{-5}\,$\AA) make them especially prone to obscuration.
The $\gamma$-rays produced near the black hole in AGN can be reprocessed into X-ray photons \citep{Inoue2020, Murase2020}, the longer wavelengths of which are less easily obscured.
Indeed, there is recent observational evidence of a potential correlation between X-rays at energies of a few tens of keV and neutrinos detected by IceCube \citep{Plavin2024, Kun2024}.
We therefore also check to see how many of our radio sources have X-ray counterparts.
To do this we use version 2 of the Millions of Optical-Radio/X-ray catalog \citep[MORX,][]{Flesch2024}, which includes probabilistic associations between VLASS sources and their likely X-ray counterparts observed by the Chandra X-ray Observatory \citep{Weisskopf2000, Weisskopf2002}, XMM-Newton \citep{Jansen2001}, the Neil Gehrels Swift Observatory \citep{Burrows2005}, and ROSAT \citep{Truemper1982}.
We find that just $7$ of our flaring radio sources ($9\,$\%) have an X-ray counterpart in MORX.

It is notable that very few of the associated radio sources have high energy electromagnetic counterparts, even though such short wavelength radiation should be a byproduct of the neutrino production.
One possible explanation for this is that the radio sources without $\gamma$-ray or X-ray counterparts are merely spurious background associations.
A second possibility, as mentioned above, is that the neutrinos are produced in a dense environment resulting in obscuration of the associated high energy photons.
A third possibility is that the radio observations are probing a larger cosmic volume (i.e., are sensitive to objects at higher redshifts) than the available $\gamma$-ray and X-ray data\textemdash we showed in Section 5.3 of \citet{Gordon2025} that VLASS variables with 4FGL counterparts tend to have lower redshifts than those without.
The observability of electromagnetic radiation is intrinsically tied to the flux density of that radiation, resulting in an observable cosmic horizon.
Neutrinos, on the other hand, only interact via the weak nuclear force and gravity, not via electromagnetism.
As such their detection is probabilistic, and subsequently neutrinos have no effective cosmic horizon.
Probing larger cosmic volumes in the electromagnetic regime therefore increases the number of opportunities to identify the sources of origin for astrophysical neutrinos.

\begin{figure}
    \centering
    \includegraphics[width=\columnwidth]{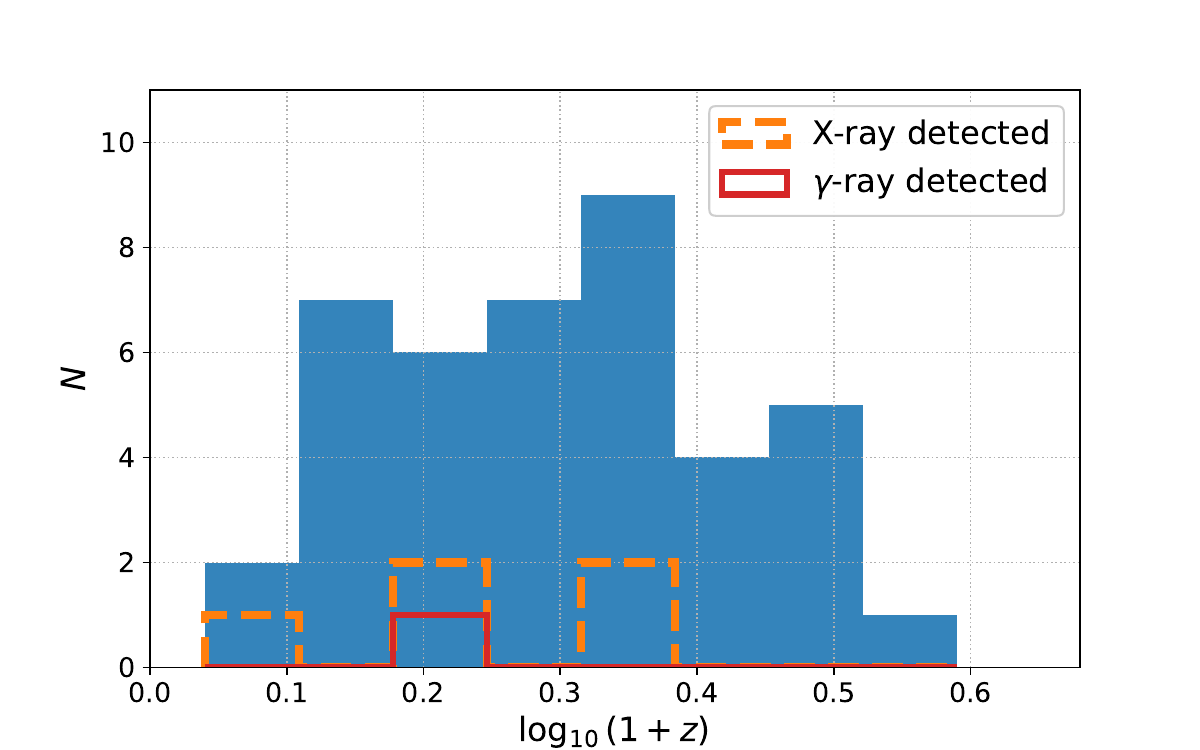}
    \caption{The distribution of redshifts for the flaring radio sources for which redshifts were found in either SDSS or LS DR8.
    The blue histogram shows the radio sources, the red solid line shows the redshift of the one source with both a Fermi detection and a redshift measurement, and the orange dashed line shows those objects in MORX with redshift measurements.}
    \label{fig:zdists}
\end{figure}

Approximately half of our flaring radio sources ($41/77$) have redshift measurements from either SDSS DR16 or \text{LS DR8}.
The redshifts lie in the range $0.17 < z < 2.42$, and have a median value of $z_{P50}=0.96$.
Only one of the sources with a $\gamma$-ray counterpart has an associated redshift ($z=0.76$), and $5$ of those with X-ray detections have redshifts ($0.23<z<1.18$), with a median value of $z_{P50}=0.67$.
We show the distribution of these sources in Figure \ref{fig:zdists}, highlighting those with X-ray and $\gamma$-ray detections by dashed orange and solid red lines, respectively. 
It is of course difficult to draw strong conclusions based on only a handful of objects, but our data are consistent with the idea that the radio flares sample a larger cosmic volume than is currently possible with $\gamma$-rays and X-rays.
Whether one or more of the possibilities presented here is responsible for the lack of high energy electromagnetic counterparts to the flaring radio sources associated with the IceCat-1 neutrino events, it is clear that our analysis using radio observations allows us to consider sources of origin for the neutrinos that would be missed in analyses limited to short wavelength observations.

\subsubsection{Infrared Colors} \label{sssec:ircolors}

\begin{figure}
    \centering
    \includegraphics[width=\columnwidth]{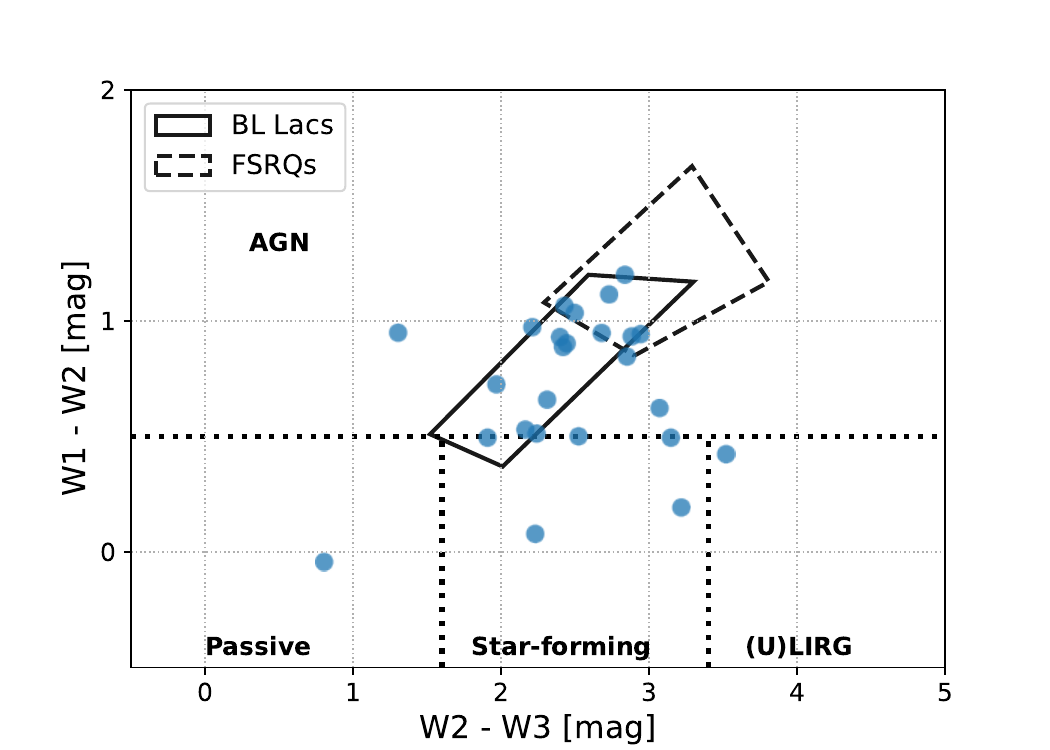}
    \caption{WISE color-color diagram for radio flares potentially associated with IceCube neutrino events.
    Blue circles show sources present in any of the flaring ($+150\,\text{day}$ to $+375\,\text{day}$ lag) samples.
    The black dotted lines show the regions occupied by \emph{passive galaxies}, \emph{star-forming galaxies}, \emph{(ultra) luminous infrared galaxies}, and \emph{AGN}, while the solid and dashed lines show the regions of IR color space where blazars are typically found.}
    \label{fig:wise-colors}
\end{figure}

Radio galaxies are often bright at infrared (IR) wavelengths, and of the $77$ flaring radio sources we are considering here, $64$ ($83\,\%$) are detected in AllWISE.
The IR colors, and in particular the brightness of an object in the three shortest wavelength WISE bands (W$1$, $3.4\,\mu$m; W$2$, $4.3\,\mu$m; and W$3$, $12\,\mu$m), of radio sources can be used to broadly characterize the source hosting the radio emission \citep[e.g.,][]{Gurkan2014, Mingo2022, Gordon2023, Wong2025}.
In Figure \ref{fig:wise-colors}, we plot a WISE color-color diagram that shows the $\text{W}1-\text{W}2$ color against the $\text{W}2-\text{W}3$ color of flaring radio sources.
Following the approach in Section 5.2 of \citet{Gordon2025}, we only include objects in Figure \ref{fig:wise-colors} where the sources are detected with at least a signal-to-noise ratio of 2 in all of the three WISE bands, and either has a measured redshift of $z<1$ or a brightness of $\text{W}1 < 14\,$mag if no redshift is available.
Objects at $z>1$ tend to move to the right (larger values of $\text{W}2-\text{W}3$) on the WISE color-color diagram as the WISE bands start to trace different regions of the galaxy spectral energy distribution, and thus may no longer be accurately classified according to diagnostic criteria developed for lower redshift galaxies \citep{Donley2012, Assef2013}.
These criteria leave us with the $25$ flaring radio sources that we show in Figure \ref{fig:wise-colors}. 

Using the diagnostic criteria of \citet{Mingo2016}, we classify the radio source as being hosted by either \emph{passive} galaxies, \emph{star-forming} galaxies (SFGs), \emph{(ultra) luminous infrared galaxies} (ULIRGs), or \emph{AGN} based on their location on Figure \ref{fig:wise-colors}.
These regions are labeled on the figure and demarcated by dotted black lines.
We also show the regions typically occupied specifically by blazars on Figure \ref{fig:wise-colors}, with the black solid line showing where BL Lacs are found, and the dashed black line highlighting the expected location of flat-spectrum radio quasars (FSRQs).
In Table \ref{tab:irpops} we show the percentage of flaring radio sources associated with each IR population, with uncertainties in these percentages calculated using a binomial approach \citep{Cameron2011}.
For reference, we also show the expected percentages in each IR population from the parent sample of radio variables in VLASS taken from \citet{Gordon2025}.

For the radio flares associated with IceCat-1 events that have the appropriate WISE data, their IR colors are not significantly different from what we would expect for any radio variable source.
This is what we should expect to see given that we have a weak signal for our excess number of radio-neutrino associations, and is perfectly consistent with the associations we do find being dominated by background sources.
Of potential note, however, is that we appear to see fewer FSRQs among our flaring sources ($\sim28\,\%$) than among the general population of variable radio sources ($\sim40\,\%$). 
Similarly, the number of SFGs ($\sim16\,\%$) appears to be slightly higher than we expect from the radio variable population ($\sim9\,\%$).
Neither of these apparent discrepancies is significant; both are at the $\sim1\sigma$ level.
However, should future analyses show a correlation between neutrinos and electromagnetic counterparts at high confidence, then significant differences may start to show between neutrino associated sources and their parent population, which may in turn help us to better understand which types of objects produce astrophysical neutrinos. 


\renewcommand{\arraystretch}{1.5}
\begin{deluxetable}{rcc}
    \tabletypesize{\scriptsize}
    \tablecaption{The IR population distributions of flaring radio sources associated with neutrino detections.
    \label{tab:irpops}}
    \tablehead{
    \colhead{IR Population} & \colhead{Radio Flares in This Paper} & \colhead{Radio Variables}\\
    \colhead{(1)} & \colhead{(2)} & \colhead{(3)}}
    \startdata
        Passive [\%] & $4_{-1}^{+8}$ & $7_{-1}^{+2}$\\
        SFGs [\%] & $16_{-5}^{+10}$ & $9_{-1}^{+2}$\\
        (U)LIRGs [\%] & $4_{-1}^{+8}$ & $1_{-0}^{+1}$\\
        AGN [\%] & $76_{-10}^{+6}$ & $83\pm2$ \vspace{0.5em}\\
        \hline
        \vspace{-1em}\\
        Blazars [\%] & $60_{-10}^{+9}$ & $68\pm3$\\
        BL Lacs [\%] & $52_{-10}^{+9}$ & $56\pm3$\\
        FSRQs [\%] & $28_{-7}^{+10}$ & $40\pm3$\vspace{0.3em}
    \enddata
    \tablecomments{The IR population based on WISE colors (1); the fraction of a population type among the radio sources associated with neutrinos in any of the flaring ($+150\,\text{day}$ to $+375\,\text{day}$ lag) samples (2); and the expected proportion of radio variables in each IR population (3) as reported in \citet{Gordon2025}.
    }
\end{deluxetable}

\subsection{Contribution of Radio Flares to the High Energy Astrophysical Neutrino Counts} \label{ssec:radiocontribution}

Assuming the excess number of associations between radio flares and neutrino events to be real and representative, we can estimate the contribution of the flaring radio source population to the observed HE astrophysical neutrinos.
Let us consider the flaring ($+150\,\text{day}$ lag) sample where we see the highest significance increase in associations between radio sources and neutrinos relative to expectations ($p=0.014$).
The number of observed associated neutrinos is $N_{\nu, \text{obs}} = 30$ and the expected number of neutrinos from random background associations is $N_{\nu, \text{bg}} = 26\pm3$.
The apparent excess number of neutrinos associated with the radio sources is therefore $N_{\nu, \text{ex}} = N_{\nu, \text{obs}}-N_{\nu, \text{bg}} = 4\pm3$.
We can compare this to the total expected number of astrophysical neutrinos expected to be detected\textemdash whether or not they are associated with a radio source\textemdash over the typical time window between radio observations in VLASS.
The median time between an Epoch 1 and Epoch 2 observation in VLASS is $974\,$days \citep{Gordon2025}.
The average number of IceCat-1 events between the start of VLASS Epoch 1 and the end of VLASS Epoch 2 was $0.072$ per day, or approximately $70$ events between VLASS observations.
Each event in IceCat-1 has a \texttt{signal} metric that is an estimate of the probability that the event was due to a \emph{real} astrophysical neutrino \citep[e.g., as opposed to an atmospheric source,][]{Abbasi2023}.
The mean \texttt{signal} for IceCat-1 events between Epoch 1 and 2 of VLASS was $0.44$, implying that of the $70$ IceCat-1 events in a $974\,$ day period, $31$ should be from an astrophysical source.
We would expect the excess number of observed neutrinos associated with radio sources to be drawn from these $31$ astrophysical events, suggesting that $13\pm10\,\%$ of astrophysical neutrinos are produced by sources experiencing an observable outburst in radio emission. 
While we use the flaring ($+150\,\text{day}$ lag) sample as an example here, we note that we see similar results using any of the flaring samples with excess radio-neutrino associations at $p<0.02$. 

While the low significance in the excess number of radio-neutrino associations prevents us from making more firm constraints on the contribution of radio flares to the observed astrophysical neutrino counts, we can compare our estimate to expectations based on previous works.
The samples of variable radio sources are selected only on their radio variability properties and are otherwise agnostic to object type.
That is to say, we do not differentiate between whether the radio emission comes from a star, galaxy, quasar, or another phenomenon.
From \citet{Gordon2025} we expect the dominant population of variable radio sources to be blazars.
Previous works have demonstrated that blazars are expected to contribute $\lesssim10\,\%$ of the observed neutrino flux \citep[e.g.,][]{Padovani2015, Aartsen2017_blazars, Murase2018}.
In Table \ref{tab:irpops} we show that $\sim 60\,\%$ of the flaring radio sources associated with neutrinos have blazar-like IR colors.
If the real radio-neutrino associations that may exist in our data are drawn randomly from the IR populations presented in Section \ref{sssec:ircolors}, then that could suggest that $\sim8\,\%$ of astrophysical neutrinos originate from blazars.
This number is remarkably similar to the finding of \citet{Kouch2025}, who estimate that $\lesssim8\,\%$ of neutrinos originate from blazars experiencing outbursts at optical wavelengths.
Given the selection bias toward blazars in the radio-variable population, a contribution of $\sim13\,\%$ to the astrophysical neutrino counts by flaring radio sources seems realistic to first order.

\subsection{Future Prospects} \label{ssec:future}


The variable radio sources used in this work were identified based solely on observations from the first two epochs of VLASS.
For the flaring samples where we see some evidence of a correlation with IceCat-1, associations between the radio source and the neutrino are limited to a single $\sim2.5\,$year window.
If the correlation we see evidence for in Section \ref{ssec:assoc-counts} is in fact real, then observing more such associations between the flares and neutrino events will reduce the shot noise and increase our confidence that a correlation exists.
Future IceCube and VLASS data may prove useful here.
VLASS Epoch 3 covers the same footprint as Epochs 1 and 2, to the same depth, and at the same cadence.
It is thus reasonable to assume that the number of radio flares between Epoch 2 and Epoch 3 will be similar to the number of flares between Epoch 1 and Epoch 2. 
VLASS Epoch 4 will observe half the footprint of Epochs $1-3$, resulting in half as many flares between Epochs 3 and 4 as between Epochs 1 and 2.
Assuming a constant average event detection rate at IceCube, the number of associations between VLASS flares and astrophysical neutrinos should scale linearly with the number of flares observed.

\begin{figure}
    \centering
    \subfigure[Predictions for IceCat-1]{\includegraphics[width=\columnwidth]{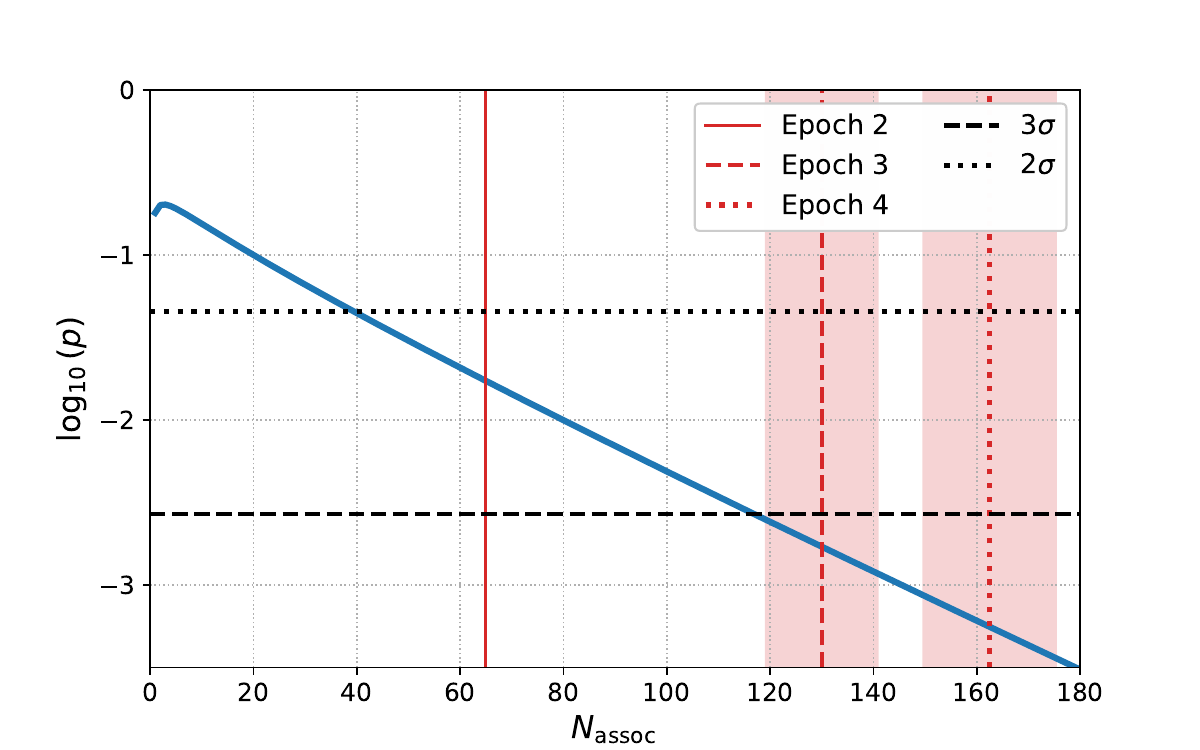}}
    \subfigure[Predictions for IceCat-2]{\includegraphics[width=\columnwidth]{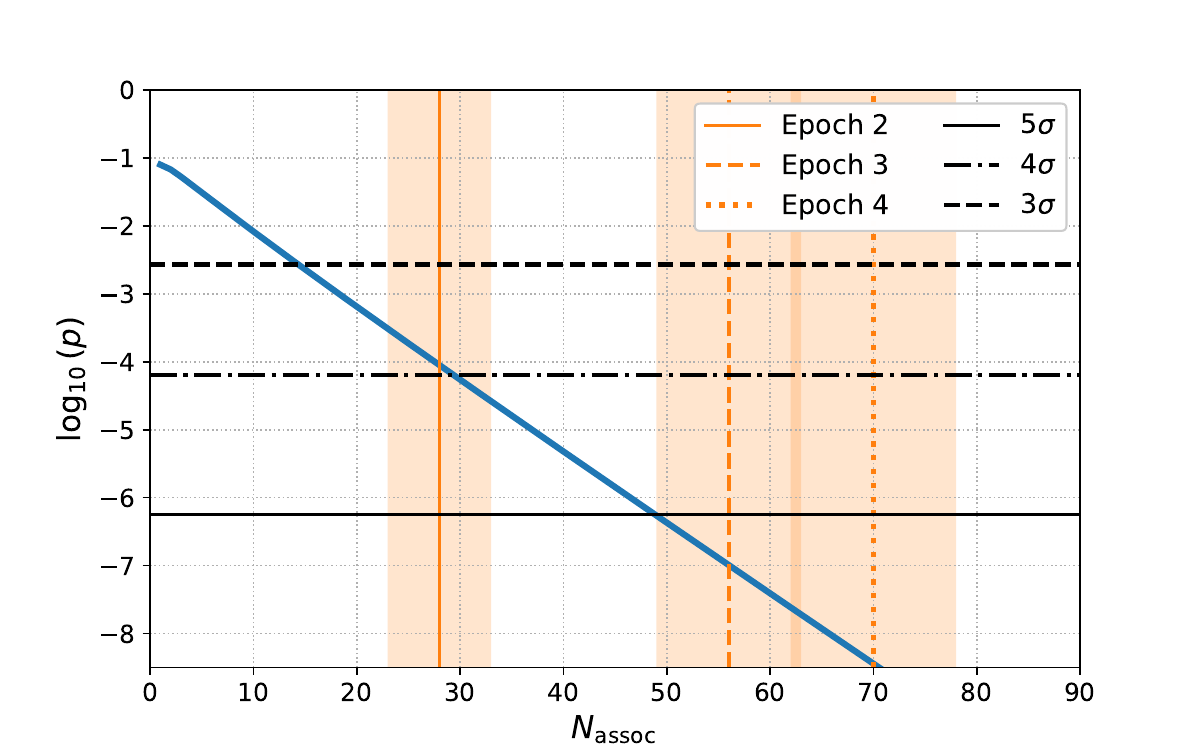}}
    \caption{Predicted $p$-value (see Equation \ref{eq:poissonpval} in text) for a given number of associations between radio flares and HE neutrinos based on the observations presented in this work and assuming a Poisson distribution.
    Panel (a) shows the predictions using IceCat-1, and panel (b) is an estimate based on the expected improvements to neutrino positional confidence in IceCat-2.
    In each panel the vertical lines show the expected number of associations at the end of a given VLASS epoch: solid vertical for the end of Epoch 2 (this work in panel a), dashed vertical for the end of Epoch 3, and dotted vertical for the end of Epoch 4. 
    The shaded regions around the vertical lines are the $\sqrt{N}$ uncertainties in the expected counts.
    The black horizontal lines highlight the $p$-values corresponding to $2\sigma$ (dotted), $3\sigma$ (dashed), $4\sigma$ (dot-dashed), and $5\sigma$ (solid) confidence.
    }
    \label{fig:prediction}
\end{figure}

We can estimate the significance for a given number of associations between neutrinos and radio flares, $N_{\text{assoc}}$ as: 
\begin{equation}
    \label{eq:poissonpval}
    p = 1-F(N_{\text{assoc}}| \lambda),
\end{equation}
where $p$ is the estimated $p$-value for a given number of associations, $F$ is the Poisson CDF, and $\lambda$ is the expected number of associations assuming the null hypothesis of no correlation between the neutrinos and radio flares.
Let us consider again the flaring ($+150\,\text{day}$ lag) sample, where an excess of radio-neutrino associations is seen at $p=0.014$, and where we observed $N_{\text{assoc}}=65$ compared to an expected $\lambda=50$.
From this we can estimate the fraction of associations that are background events as $f_{\text{bg}}=\lambda/N_{\text{assoc}} = 0.77$.
In Figure \ref{fig:prediction}a we show the predicted $p$-value (Equation \ref{eq:poissonpval}) for increasing numbers of associations between VLASS flares and IceCat-1 neutrinos assuming this value for $f_{\text{bg}}$, with the expected number of associations by the end of VLASS Epochs 3 and 4 highlighted by red vertical dashed and dotted lines, respectively.
If there is actually a correlation between VLASS flares and IceCat-1 neutrinos at the level we see in this work, then we would expect to detect that correlation with $>3\sigma$ confidence by incorporating the VLASS Epoch 3 data.

A future version of IceCat known as IceCat-2 will include improved angular resolution for both new and historical neutrino events \citep{Zegarelli2025}.
The positional uncertainties in IceCat-2 are expected to cover $\sim1/4$ of the solid angle subtended by the IceCat-1 uncertainties, leading to a corresponding reduction in the number of random background associations with electromagnetic counterpart candidates.
From our flaring ($+150\,\text{day}$ lag) sample, the number of excess associations over the expected random background can be estimated as $N_{\text{ex}} = 65-50 = 15$.
With the improved IceCat-2 uncertainties we would expect $1/4$ of the background associations that are observed in IceCat-1, leading to $\sim28$ associations ($N_{\text{ex}} + \lambda/4$), with $p = 1-F(28| 13) = 9\times10^{-5}$.
In panel (b) of Figure \ref{fig:prediction}, we show the expected $p$-values for increasing numbers of associations between radio flares and neutrinos assuming that the IceCat-2 data does in fact reduce the shot noise as expected.
By the end of VLASS Epoch 3 we anticipate $N_{\text{assoc}}=56$ and $\lambda = 26$ using the IceCat-2 data, an observation that
would have $p\sim10^{-7}$, or $>5\sigma$ confidence.
While we have used the $+150\,$day lag sample as being representative of our results here, we note that we obtain similar predictions when estimating $N_{\text{assoc}}$, $\lambda$, and $N_{\text{bg}}$ using the results from any of our samples where we see an excess of radio-neutrino associations at $p<0.02$.
Of course, as mentioned in Section \ref{ssec:assoc-counts}, these are not independent samples. 
However, should these data be broadly representative of the scale of the effect we observe in this work, then future VLASS and IceCube data should be able to provide tighter constraints on the contribution of radio flares to the population of HE neutrinos than previously possible.

\section{Summary and Conclusions} \label{sec:summary}

In this work we investigate whether variable radio sources contribute to the astrophysical neutrinos observed by IceCube.
Radio-variable objects are widely considered to be strong candidates for neutrino sources, but until recently such research has generally been limited to the brightest and most well-studied variable radio sources.
The recent advent of wide-area, deep (flux densities down to a few mJy), time-domain radio surveys like VLASS allows for the expansion of radio-neutrino cross-association stacking experiments to include a more complete sampling of the radio-variable population.
We use data from the first two epochs of VLASS and the IceCat-1 catalog of neutrino events to perform such an analysis.
Our key findings are summarized below.
\begin{itemize}
    \item We find an excess number of associations between flaring radio sources and neutrinos that are both spatially \emph{and} temporally associated at $>2\sigma$ confidence.
    \item Should this apparent correlation prove to be real, we see evidence that the neutrino event may precede the radio outburst by hundreds of days, similar to previously observed delays between $\gamma$-ray and radio emission in blazar outbursts.
    \item Radio flares that are spatially and temporally associated with neutrinos are typically located closer on the sky to the neutrino event than simulated random associations. 
    Real associations have a median angular separation of $140^{\prime}$ compared to $180^{\prime}$ for random associations.
    \item The majority ($\sim90\,\%$) of radio sources associated with IceCat-1 events are not detected at $\gamma$-ray or X-ray wavelengths, highlighting the importance of radio time-domain surveys in identifying candidate electromagnetic counterparts to astrophysical neutrinos.
    \item The excess number of neutrinos identified by these associations is consistent with flaring radio sources contributing $\sim13\,\%$ of the astrophysical neutrinos observed by IceCube, in agreement with previous studies.
    \item The number of associations between variable radio sources and neutrinos is consistent with the expected background when no time correlation is required, i.e., there is no evidence of a spatial-only correlation between radio variables and neutrinos.
    \item Assuming our results are representative, we predict that the significance of the excess number of associations between radio flares and neutrinos will exceed $3\sigma$ confidence by the end of VLASS Epoch 3.
    Should the improved uncertainties in the sky position of neutrino events in the forthcoming IceCat-2 also lower the shot noise as expected, then we anticipate $>5\sigma$ confidence in the excess number of associations between radio flares and HE astrophysical neutrinos.
\end{itemize}

\section*{Acknowledgments}

The authors thank the anonymous referees for their helpful comments that have strengthened this paper.

In carrying out this work we made use of the following software packages and tools:
AstroPy \citep{Astropy2013, Astropy2018, Astropy2022},
Matplotlib \citep{Hunter2007},
NumPy \citep{Harris2020},
and TOPCAT \citep{Taylor2005}.

The VLA is operated by NRAO, a facility of the National Science Foundation operated under cooperative agreement by Associated Universities, Inc.

WISE is a joint project of the University of California, Los Angeles, and the Jet Propulsion Laboratory/California Institute of Technology, and NEOWISE, which is a project of the Jet Propulsion Laboratory/California Institute of Technology. WISE and NEOWISE are funded by the National Aeronautics and Space Administration.


\bibliographystyle{aasjournal.bst}

\bibliography{vlassxicecube}

\newpage
\begin{appendix}

\section{Supplementary Data}
\label{ap:assocdata}

As part of this work, we have constructed a catalog of flaring radio sources from VLASS that are spatially and temporally consistent with HE neutrino detections from IceCube. 
We make this catalog available to the wider scientific community as an electronic data table available at \url{https://zenodo.org/records/23223067}.
In Table \ref{tab:radio-nu-assocs} we show an example of the structure of the catalog, which lists the IDs of the radio source and neutrino event, their observation dates, and the brightness of the radio source in Epochs 1 and 2.
The first 30 rows are shown in Table \ref{tab:radio-nu-assocs}. The full table has $104$ rows containing $99$ distinct radio sources and $46$ different neutrino events.

\renewcommand{\arraystretch}{1.5}
\begin{deluxetable}{ccccccccc}
    \tabletypesize{\footnotesize}
    \tablecaption{Associations of flaring radio sources from VLASS with high energy neutrinos from IceCat-1 \label{tab:radio-nu-assocs}}
    \tablehead{
    \colhead{VLASS1QLCIR} & \colhead{$S_{\text{E1}}$} & \colhead{e\_$S_{\text{E1}}$} & \colhead{MJD$_{\text{E1}}$} & \colhead{$S_{\text{E2}}$} & \colhead{e\_$S_{\text{E2}}$} & \colhead{MJD$_{\text{E2}}$} & \colhead{NAME} & \colhead{MJD$_{\nu}$}\\
    \colhead{} & \colhead{[mJy]} & \colhead{[mJy]} & \colhead{} &  \colhead{[mJy]} & \colhead{[mJy]} & \colhead{} & \colhead{} & \colhead{}\\
    \colhead{(1)} & \colhead{(2)} & \colhead{(3)} & \colhead{(4)} &  \colhead{(5)} & \colhead{(6)} & \colhead{(7)} & \colhead{(8)} & \colhead{(9)}}
    \startdata
    J003026.30$+$811216.1 & 8.70 & 0.18 & 58074.1352 & 16.49 & 0.25 & 59104.3904 & IC190629A & 58663.8085\\
    J013751.09$+$092319.1 & 112.28 & 0.32 & 58051.3594 & 195.32 & 0.50 & 59079.3640 & IC181120A & 58442.7087\\
    J013821.05$-$001126.9 & 7.27 & 0.24 & 58103.1805 & 14.85 & 0.32 & 59076.4638 & IC191122A & 58809.9480\\
    J014459.23$+$125949.1 & 32.53 & 0.22 & 58033.2624 & 90.30 & 0.33 & 59063.4126 & IC181120A & 58442.7087\\
    J020344.25$+$304237.9 & 156.35 & 0.93 & 58580.7123 & 237.37 & 0.93 & 59506.3819 & IC200614A & 59014.5287\\
    J021842.20$+$314926.8 & 39.70 & 0.29 & 58568.7627 & 88.86 & 0.23 & 59515.3644 & IC200614A & 59014.5287\\
    J022601.54$+$135726.9 & 12.41 & 0.21 & 58031.4515 & 20.07 & 0.28 & 59062.4491 & IC180613A & 58282.9824\\
    J024037.87$+$125017.4 & 5.63 & 0.23 & 58123.0212 & 10.29 & 0.35 & 59063.5086 & IC170824A & 57989.5538\\
    J024037.87$+$125017.4 & 5.63 & 0.23 & 58123.0212 & 10.29 & 0.35 & 59063.5086 & IC180613A & 58282.9824\\
    J024712.77$+$201042.1 & 21.18 & 0.22 & 58641.5817 & 41.03 & 0.22 & 59501.4683 & IC191231A & 58848.4584\\
    J025859.64$+$003618.1 & 19.11 & 0.17 & 58077.2998 & 31.37 & 0.29 & 59099.3719 & IC210717A & 59412.6562\\
    J030416.07$+$232618.6 & 6.57 & 0.20 & 58566.7861 & 12.08 & 0.21 & 59543.3777 & IC191231A & 58848.4584\\
    J030416.07$+$232618.6 & 6.57 & 0.20 & 58566.7861 & 12.08 & 0.21 & 59543.3777 & IC211125A & 59543.2655\\
    J030649.32$+$175040.2 & 3.37 & 0.27 & 58566.8811 & 6.48 & 0.21 & 59525.4141 & IC191231A & 58848.4584\\
    J030958.43$+$191229.1 & 12.82 & 0.28 & 58566.8610 & 21.11 & 0.20 & 59525.3801 & IC191231A & 58848.4584\\
    J031118.18$+$233234.7 & 6.31 & 0.27 & 58566.7859 & 17.43 & 0.22 & 59543.3781 & IC191231A & 58848.4584\\
    J031805.36$+$223235.1 & 7.36 & 0.22 & 58566.7999 & 23.34 & 0.22 & 59543.3593 & IC191231A & 58848.4584\\
    J031845.65$+$213436.7 & 40.43 & 0.46 & 58566.8190 & 63.38 & 0.32 & 59543.3371 & IC191231A & 58848.4584\\
    J031928.61$+$382644.3 & 10.06 & 0.21 & 58586.7115 & 15.93 & 0.20 & 59546.3305 & IC200911A & 59103.5971\\
    J032356.26$+$361211.1 & 15.39 & 0.23 & 58586.7496 & 25.25 & 0.22 & 59546.3689 & IC200911A & 59103.5971\\
    J032447.74$+$370944.5 & 9.69 & 0.21 & 58586.7320 & 17.70 & 0.21 & 59546.3512 & IC200911A & 59103.5971\\
    J033104.48$+$394442.2 & 6.67 & 0.25 & 58586.6989 & 12.84 & 0.20 & 59546.3177 & IC200911A & 59103.5971\\
    J043515.21$+$054810.7 & 19.94 & 0.24 & 58063.2452 & 32.49 & 0.30 & 59078.6202 & IC180213A & 58162.3778\\
    J044624.72$+$110521.9 & 13.77 & 0.23 & 58084.1751 & 24.66 & 0.28 & 59111.6387 & IC190712A & 58676.0523\\
    J050404.07$+$141737.3 & 10.83 & 0.23 & 58029.3978 & 16.63 & 0.28 & 59062.5928 & IC190712A & 58676.0523\\
    J050618.18$+$144217.0 & 9.41 & 0.23 & 58029.3933 & 18.28 & 0.28 & 59062.5883 & IC190712A & 58676.0523\\
    J050625.00$+$080939.5 & 13.28 & 0.22 & 58084.2243 & 28.10 & 0.25 & 59111.5844 & IC180123A & 58141.6771\\
    J052140.81$+$010255.6 & 19.34 & 0.31 & 58117.2904 & 36.53 & 0.35 & 59097.5007 & IC190317A & 58559.8319\\
    J054415.93$+$023831.5 & 21.96 & 0.24 & 58117.3078 & 32.98 & 0.28 & 59097.5200 & IC190317A & 58559.8319\\
    J054640.41$+$073749.1 & 8.60 & 0.25 & 58028.4471 & 13.29 & 0.31 & 59070.6439 & IC200421A & 58960.0246\\
    ... & ... & ... & ... & ... & ... & ... & ... & ...\\
    ... & ... & ... & ... & ... & ... & ... & ... & ...\\
    ... & ... & ... & ... & ... & ... & ... & ... & ...
    \enddata
    \tablecomments{The ID of the VLASS source from the Epoch 1 catalog (1); the VLASS flux density in Epoch 1 (2); the uncertainty in $S_{\text{E1}}$ (3); the MJD of the VLASS Epoch 1 observation (4); the VLASS flux density in Epoch 2 (5); the uncertainty in $S_{\text{E2}}$ (6); the MJD of the VLASS Epoch 2 observation (7); the name of the IceCat-1 event (8); and the MJD of the neutrino detection (9).
    The first 30 rows are shown here with the full table of 104 rows available online at \url{https://zenodo.org/records/23223067}.
    }
\end{deluxetable}

\end{appendix}

\end{document}